\documentclass[12pt]{article}
\usepackage{jheppub}

\usepackage{amsmath,epsfig}
\usepackage{amssymb,amsfonts}
\usepackage{latexsym}
\usepackage[latin1]{inputenc}
\usepackage{subeqnarray}
\usepackage{array}
\usepackage{xcolor}

\usepackage{graphicx}
\usepackage{subfigure}
\usepackage{longtable}
\usepackage{mathdots}
\usepackage{physics}
\usepackage{enumitem}
\usepackage{oubraces}
\usepackage{cancel}

\relax
\renewcommand{\theequation}{\arabic{section}.\arabic{equation}}
\def\be{\begin{equation}}
\def\ee{\end{equation}}
\def\bea{\begin{eqnarray}}
\def\eea{\end{eqnarray}}
\newcommand\fverb{\setbox\pippobox=\hbox\bgroup\verb}
\newcommand\fverbdo{\egroup\medskip\noindent%
                        \fbox{\unhbox\pippobox}\ }
\newcommand\fverbit{\egroup\item[\fbox{\unhbox\pippobox}]}
\newcommand{\ha}{{1 \over 2}}

\newcommand{\bear}{\begin{eqnarray}}

\newcommand{\eear}{\end{eqnarray}}
\newcommand{\de}{\partial}
\newcommand{\bsea}{\begin{subeqnarray}}
\newcommand{\esea}{\end{subeqnarray}}
\newbox\pippobox

\let\oldsqrt\sqrt
\def\sqrt{\mathpalette\DHLhksqrt}
\def\DHLhksqrt#1#2{%
\setbox0=\hbox{$#1\oldsqrt{#2\,}$}\dimen0=\ht0
\advance\dimen0-0.2\ht0
\setbox2=\hbox{\vrule height\ht0 depth -\dimen0}%
{\box0\lower0.4pt\box2}}

\newcommand{\eql}[2]
{ \begin{equation} \label{#1}
 #2
\end{equation}}

\def\d{\delta}
\def\g{\gamma}

\def\6{\partial}

\def\a{\alpha}

\def\le{\left}
\def\ri{\right}

\def\cO{{\cal O}}

\def\e{\epsilon}
\def\m{\mu}
\def\n{\nu}

\def\s{\sigma}
\def\t{\T}
\def\sp{\;\;\;,\;\;\;}

\def\p{\partial}

\def\sq
\def\a{\alpha}
\def\b{\beta}

\def\hri#1#2{\href{http://arxiv.org/abs/#1}{[ArXiv:#1]#2}}
\def\hre#1#2{\href{http://arxiv.org/abs/#1/#2}{[ArXiv:#1/#2]}}

\def\e{\epsilon}

\def\d{\delta}

\def\L{\Lambda}

\def\D{\Delta}

\def\dd{{\rm d}}
\def\AA{{\cal A}}

\def\DD{{\cal D}}

\def\FF{{\cal F}}
\def\GG{{\cal G}}

\def\MM{{\cal M}}

\def\RR{{\cal R}}
\def\SS{{\cal S}}

\def\WW{{\cal W}}

\def\Vol{{\sqrt{-g}}}
\def\WN{{\widetilde\nabla}}
\def\WG{{\widetilde\Gamma}}
\def\mn{{\m\n}}

\def\emn{{\eta_{\m\n}}}

\def\nd{{\text{and}}}
\def\T{{\Theta}}
\def\gk{{\mathfrak{g}}}

\title{On multi-field flows in gravity and holography}

\author{Francesco Nitti, Leandro Silva Pimenta, Dani\`ele A.~Steer}

\affiliation{\href{http://www.apc.univ-paris7.fr}{APC, AstroParticule et Cosmologie}, Universit\'e Paris Diderot, CNRS/IN2P3, CEA/IRFU, \\
Observatoire de Paris, Sorbonne Paris Cit\'e,\\
 10, rue Alice Domon et L\'eonie Duquet, 75205 Paris
Cedex 13, France}

\abstract{We perform a systematic analysis of flow-like solutions in
theories of Einstein gravity coupled to multiple scalar fields, which
arise as holographic RG flows as well as in the context of
cosmological solutions driven by scalars.  We use the first order
formalism and the superpotential formulation to classify solutions
close to generic extrema of the scalar potential, and close to
``bounces,'' where the flow is inverted in some or all directions and
the superpotential becomes multi-valued. Although the superpotential
formulation contains a large redundancy, we show how this can be
completely lift by suitable regularity conditions. We place the first
order formalism in the context of Hamilton-Jacobi theory, where we
discuss the possibility of non-gradient flows and their connection to
non-separable solutions of the Hamilton-Jacobi equation. We argue that
non-gradient flows may be useful in the presence of global symmetries
in the scalar sector.

}

\begin{document}
\maketitle


\section{Introduction and summary}
\label{s:intro}
Theories of gravity coupled to scalar fields are an important
ingredient in many subjects including cosmology (inflation), large
distance gravitational physics  (models of
dark energy,  modified gravity theories) and high energy theory and
phenomenology (string theory, supergravity,  brane world models), as
well as in the context of the  gauge/gravity duality. The presence of multiple
scalars often makes the physics qualitatively different from the 
single scalar case. The latter is relatively
manageable when looking for simple solutions in which the fields depend
on only one coordinate (as in the case of cosmology or simple holographic
renormalisation group flows), since  one can use  the  scalar field
as the coordinate which describes the evolution.  That said, in many cases, having multiple scalars  with
non-trivial evolution  is inevitable, and the
reduction to a  single-field may be too simplistic.  This is the case for
example in  cosmology and in gauge/gravity duality, the latter of which will be the main focus of this work. 

The gauge/gravity duality is the conjectured equivalence between
a large $N$ gauge theory in $d$-dimensional flat space-time ({\em
  boundary theory}), and
a higher dimensional  theory with dynamical gravity in higher
dimensional curved space-time ({\em bulk theory}), \cite{Maldacena97,Gubser98,Witten98}. In this context
bulk scalar fields correspond to couplings of single-trace scalar operators in
the dual, boundary field theory. Evolution in the bulk geometry
corresponds to evolution under the renormalisation group (RG) in the
boundary theory.  Solutions of the bulk equations in which the
scalars run along the holographic coordinate (which parametrise a non-compact
direction among  those which
are {\em extra} with respect to the boundary coordinates) are called  
{\em holographic RG flow} solutions
\cite{9810126,9903190,9904017,deboer,Bianchi:2001de,HaroSkenderisSolodukhin,skenderisham}.

Holographic RG flows have been widely studied, and often (especially
in phenomenological models)  it is assumed for simplicity that only one of the scalars runs.
However, from the field theory point of view, it is
clear that this is an oversimplification: any QFT has an infinite number
of operators, which will generically mix under the RG flow. Many operators  will
start running even when the corresponding couplings  are not turned on in
the  far UV.  Therefore, even though in some cases one can, to a first
approximation, hope to neglect this mixing, in general this will not
be possible and one should consider solutions with
multiple scalars evolving.   

The need to consider multifield scenarios also arises in a cosmological setting, for instance inflation, see
e.g.~\cite{Langlois:2008wt,Langlois:2008qf,Renaux-Petel:2015mga}.  This is because in many cases, the truncation to a single field
fails to capture some important aspects of the full dynamics. Similar issues arise in the context of supergravity truncations, see e.g.~\cite{Celi:2004st}.  

In this work, we consider   $d+1$-dimensional of Einstein gravity
coupled to $N$ scalar fields $\phi^r$ with a generic scalar potential,
and we focus on {\em flow} solutions,  which  depend on a single
coordinate $u$ (space-like in holography and time-like in
cosmology) and which can be brought to the general form
\be\label{intro0}
ds^2 = du^2 + e^{A(u)}\eta_{\mu\nu}dx^\mu dx^\nu, \qquad \phi^r
  = \phi^r(u), \qquad r=1,\ldots, N. 
\ee
In holography these solutions describe Poincar\'e-invariant vacuum (or false vacuum)
states of the dual $d$-dimensional field theory; in cosmology they
represent flat FRW space-times.  Our analysis will be
mostly framed in the language of holographic RG flows,  but many of
our results carry over unchanged to more general contexts.

When studying solutions of the form (\ref{intro0})  it has often been very
useful to rewrite Einstein's equations as first order flow equations,
governed by a {\em superpotential} $W(\{\phi^r\})$
\cite{deboer,Bianchi:2001de,Skenderis:1999mm}, i.e. a function
$W(\phi)$ on the scalar field manifold such that the ansatz (\ref{intro0}) solves the equations 
\be
{d\phi^r \over d u} = \GG^{rs}{\partial W \over \partial \phi^s}, \qquad {d A \over d
  u} = -2(d-1) W ,  
  \label{intro1}
\ee  
where $\GG^{rs}$ is the (inverse) metric on field space.  The
superpotential  is determined by a
 coordinate-invariant differential equation
given schematically by
\be \label{intro2}
\GG^{rs}\de_r W \de_s W - W^2 = V,  
\ee
where $V$ is the scalar field potential. 
The superpotential formulation is a
way of grouping together solutions  into classes which share the same
geometric features. Solutions in the same class differ by the initial
condition of the flow equations \eqref{intro0}.  This way of organising the space of solutions also has
applications in cosmology, as argued in \cite{Binetruy:2014zya}.



In holography, the superpotential equation in the single field case
has been widely studied and in the single-field case the qualitative
features of the solutions are known for general potentials
\cite{PapaMulti,1106.4826,Martelli:2001tu,wenliang,Bourdier,multibranch},
The first order formalism is a very convenient way of classifying
solutions close to an extremum of the potential (IR or UV fixed
point)  and when the scalar runs to infinity \cite{0707.1349}; it
determines the holographic $\beta$-function for the coupling dual to
the bulk scalar, by $\beta(\phi) = -  \de_\phi \log W $
\cite{deboer,wenliang}; it provides a $c$-function which interpolates
monotonically between the UV and IR central charges
\cite{9904017,Bianchi:2001de,1006.1263} ; it makes it very simple to
write counterterms for holographic renormalisation
\cite{Bianchi:2001kw,1106.4826} and the gravitational on-shell action
\cite{wenliang}.  The first order formalism  also allows to uncover and classify ``exotic'' features which
cannot occur in perturbative field theory, such as inversion of  the
direction of  a holographic RG flow, at which points the
superpotential becomes multi-valued, as it was  
  observed early on in  \cite{Sonner:2007cp,Gursoy:2008za} and
  discussed in detail in 
 \cite{multibranch}. 

The multi-field case is much more involved, and a systematic analysis has so
far been missing. The main reason is that the multi-field
superpotential equation is a non-linear partial differential equation
(as opposed to an ordinary differential equation in the single-field
case). As a consequence, several properties which are automatic in the
simple-field case (e.g. the very existence of a superpotential, the gradient
property  of all flows which we discuss below), are less obvious when
many scalars are involved. 

Another major difference is that, in the single-field case,  the number of
integration constants in the first order formulation is {\em exactly}
the same as in the standard second order form of Einstein's
equation. This implies that there is a one-to-one correspondence
between solutions of Einstein equations and solutions of the flow
equations, and given a solution of the form (\ref{intro0})  there always  exists a unique
superpotential which can be reconstructed following an algorithmic
procedure. Instead,  with  multiple scalars,  the correspondence between first order and second order
formulation is many to one: the same solution of the Einstein-scalar
system can arise from {\em many} different superpotentials, as  we will
see in explicit examples. 

This ambiguity goes beyond the usual   dependence of the
field theory $\beta$-functions on field
redefinitions. In holography, once the boundary conditions in the UV
are fixed, different superpotentials (different
$\beta$-functions) correspond to different vacua of the same
theory.  For a single field, there is a one-to-one correspondence
between vacua and $\beta$-functions, whereas  in the multi-field
case  there seem to be many more $\beta$-functions  to describe
the same set of physical solutions (flows). 
One of our results will be to show that this degeneracy is lifted by
appropriate   regularity conditions in the IR, which can eliminate
all but one (or a discrete set) of beta-functions.

In this work we pursue two main goals. 
On the one hand, we  perform a general analysis of the space of solutions in terms of first order flows
(\ref{intro0}).
On the other hand, we will analyse the system from
the point of view of Hamilton-Jacobi theory and investigate if and how  
the existence of a  superpotential is justified, and more generally 
when one can describe the space of solutions using only gradient
flows. 

In the first part of the paper we analyse the space of solutions of
the superpotential equation and the corresponding flows,  extending to the
multi-field case the systematic analysis that was performed in
\cite{multibranch}. In particular we give a general classification of
the behavior at singular points, where 
 some or all partial derivatives of
 $W$ vanish. These may be  of two types: 
\begin{enumerate}
\item Extrema of the potential: these may be reached in the far
  UV or the far IR, and they are the endpoints of the flows.
\item Points which are not extrema of the potential: here, 
  one or more directions of the flow are inverted, and $W$  becomes
  multi-branched. These are not endpoints, as the solution can be
  continued smoothly past these points. They were referred to as {\em
    bounces} in \cite{multibranch}
\end{enumerate}

Unlike in the case of a single field, a generic extremum of the potential can play the role  both of an
IR and an UV fixed point, depending on which directions in field space
are running\footnote{With the exception of a local maximum, which can
  only be an UV fixed point.}.
Close to
the extremum, the behaviour of the superpotential is a
generalisation of the single field case:  a
 analytic part which is universal (up to discrete choices),  plus a sub-leading part which contains  the
integration constants. The new feature in the multi-field case is, on
the one hand, that we are now in the presence of integration {\em
  functions}, therefore the description is highly redundant. We will 
identify a
restricted class of solutions which contain
just enough integration constants to provide all possible solutions of
Einstein's equations of the form \eqref{intro0}.

Furthermore, we find that imposing  an appropriate  regularity condition 
around {\em minima} of the scalar potential  lifts all
continuous deformations, 
leaves only one  physical
vacuum  (or at most 
a discrete set of 
), and eliminates the redundancy in the first order description {\em while still
allowing a continuous choice of the UV source parameters.} This last
requirement is crucial  because we  do not want to restrict
the values of the UV couplings, which enter as initial conditions of
the flow around a maximum of the potential.  Indeed, in  a QFT it
should be possible to change the couplings continuously, at least in a
certain range, without making the theory inconsistent\footnote{This
  does not mean that any value of the coupling is allowed, as there
  could be regions in the space of couplings which are forbidden. This
is true even in perturbative QFT, for example we should require
$\lambda>0$ in $\lambda \varphi^4$ theory, or $g^2>0$ in Yang-Mills
theory. However the allowed regions must for continuous deformations.}.

As in the single-field case, away from extrema of the potential a
solution to the  superpotential equation can {\em bounce} and become
multi-branched at certain special points. When this occurs,  the flow of
one or more scalars inverts its direction, causing a breakdown of the
first-order formalism.
In  the multi-field case  the
structure of bounces is much richer than for a single field. First,
bounces can now occur on a hyper-surface of any dimension up to $N-1$
of the scalar manifold. Second, there are two qualitatively different
kinds of bounces:
\begin{itemize}
\item Complete bounces, where {\em all} the scalars change direction
  at the same time
\item Partial bounces, where only some of the scalars invert their
  flow. 
\end{itemize}
Complete bounces occur on sub-manifold lying on
 equipotential hyper-surfaces
 whereas partial bounces can occur anywhere in field
space except at extrema of $V(\phi)$. Close to a bounce the superpotential has several branches, and
we show how to glue them together so that the flow is smooth. 
Interestingly, close to a complete bounce, 
the superpotential equation takes the same form as the Eikonal equation for
geometric optics close to a surface with vanishing index of
refraction. Thus, complete bounces are analogous to the phenomenon of
total internal refraction which gives rise to mirages. 

As we have mentioned, in the multi-field case the existence of a
superpotential, for which the flow has the gradient form
(\ref{intro1}) is not obvious. The second part of this work explores  the questions 1) whether
a superpotential description is always possible 2) whether 
more general descriptions of the space of solutions in terms of
non-gradient flows may sometimes be useful.  

 The appropriate framework to answer these
questions  is Hamilton-Jacobi theory, whose connection with the first order form
in holography and cosmology is well known, \cite{deboer,0404176,SkenderisTownsend,PapaCan,PapaLectures}. In Hamilton-Jacobi
theory,  one can always find a first order gradient flow description in the {\em extended}
  $N+1$ dimensional field space with coordinates $(A,\phi^r)$,
  generated by Hamilton's principal function $\SS(A,\phi^r)$. However,
  as we will see, in order to have an $N$-dimensional gradient flow on
  the scalar manifold parametrised by $\{\phi^r\}$,  $\SS(A,\phi^r)$
  must have a specific separable form \cite{Celi:2004st}. Gradient flows in extended
  field space, with a non-separable principal function, have  been
  shown to arise   in connection with black hole solutions
  \cite{Lindgren:2015lia}. 

As we will discuss, the answer to both  questions raised above is
positive. The superpotential description arises from a
special class of {\em separable}  Hamilton-Jacobi (HJ)
principal functions. Locally, any solution of Einstein's equation can be seen as
arising from such a separable  HJ function, which implies that a
superpotential can always be found locally in field space. Although it
may not be globally defined as a smooth function, we can use the results of the first part of
this work to glue together consistently solutions in different
regions. 
 Furthermore, in  the case of holography, 
separable solutions contain already enough integration constants to
describe all possible RG flows starting from a UV maximum. 

A situation when it may be useful {\em not} to use a separable
function is in the presence of  global symmetries: in this case, we
find that if we want to classify solutions in terms of the value of
the corresponding conserved charges,  a gradient flow
description is impossible, and a superpotential cannot be
defined. Classifying solutions according to the value  of conserved
charges may be a useful option in multi-scalar cosmology
 as it may simplify the treatment.

If we restrict ourselves to holographic RG flows however, one can
argue that  all conserved charges must vanish,  and non-gradient flows
an unnecessary.  The reason is that, in holography,
bulk symmetries should be gauged, as  they also imply global
symmetries of the boundary theory at the fixed point:  there should
be bulk gauge fields 
corresponding to the boundary conserved currents. As a consequence, 
  in the absence of non-trivial  gauge fields in the
solution,  the value of all conserved quantities must be 
zero by gauge invariance (or, equivalently, by Gauss's law).

This paper is organised as follows. 

In Section \ref{s:W} we lay out our
setup and introduce the first order formalism. 

In Section \ref{s:hol} we
discuss holographic RG flows in terms of the superpotential. We first
classify all possible solutions around generic extrema of the
scalar potential, their continuous parameters, and we differentiate
between UV and IR solution, in subsection \ref{ss:nex}. Then,  in
subsection \ref{ss:bou}, we turn to the analysis of multi-branched
solutions. 

In Section \ref{s:can} we make the connection with Hamilton-Jacobi
theory. In subsection \ref{ss:grad}  we relate non-gradient flows to
non-separable solutions of the Hamilton-Jacobi equation. We discuss
global symmetries in subsection \ref{ss:sym}, and in subsection
\ref{ss:gauge} we discuss the effect of the gauging of such
symmetries. 

Finally in Section \ref{s:end} we summarise our conclusion and propose
further directions. 

Some of the more technical details of our calculations, as well as a
review of Hamilton-Jacobi theory, are left to the Appendix. 
 

\section{Setup}
\label{s:W}

\subsection{Action, field  equations and vacuum ansatz}

Our starting point is the Klein-Gordon action for $N$ self-interacting scalar fields minimally coupled to gravity, in $d+1$ dimensions, 
\begin{subequations}\label{gen0}
\begin{align}
&S=
M^{d-1}
\int_{\MM} \dd^{d+1}x \Vol \left[
R
-\ha \GG_{rs}\partial_a \phi^r\partial^a\phi^s -V(\phi^r) \right]+S_{GH}
\label{action}
~,\\
&S_{GH}=-2M^{d-1} \int_{\partial \MM} \dd^dx \sqrt{h} K \,.
\end{align}
\end{subequations}
Here $a=0,...,d$ and early alphabet letters are space-time indexes, $r=1,...,N$ and middle alphabet letters are field-space indexes. The field space metric $\GG_{rs}(\phi^1,\ldots,\phi^N)$ is assumed positive-definite and non-degenerate. In the Gibbon-Hawking term, $S_{GH}$, $h_{ab}$ is, as usual, the induced metric on the space-time boundary $\partial \MM$ which has extrinsic curvature $K_{ab}$. We will consider solutions preserving $d$-dimensional Poincar\'e invariance,
\be
\label{gen1}
\dd s^2={\dd u^2}+e^{2A(u)}\eta_\mn\dd x^\m \dd x^\n~, ~~
\phi^s=\phi^s(u)
~.\ee
In the gauge/gravity duality, these solutions correspond to RG flows in a
space of $N$ coupling constants, each corresponding to one scalar operator. The same ansatz \eqref{gen1} can be applied to cosmology after two Wick rotations.

The Klein-Gordon and Einstein equations are
\begin{subequations}\label{gen3}
\begin{align}
&
\ddot\phi^r
	+ \WG^r_{pq}\dot\phi^p\dot\phi^q
+d \dot A\dot \phi^r
	-\GG^{rp}{\p V\over \p\phi^p}=0
~,
\label{gen3a}
\\
&
d(d-1)\dot A^2-\ha \GG_{rs}\dot\phi^r\dot\phi^s+ V(\phi)=0
~,
\label{gen3b}
\\
&
2(d-1)\ddot A +\GG_{rs}{\dot \phi^r}{\dot \phi^s}=0
~,
\label{gen3c}
\end{align}
\end{subequations}
where $\cdot={\rm d}/{\rm d}u$, and
\eql{}{
	\WG^r_{pq}
	=\ha \GG^{rs}\le(
	\p_p\GG_{sq}
	+\p_q\GG_{sp}
	-\p_s\GG_{pq}
	\ri)
	~,
	\nonumber
}
with $\partial_p = \frac{\partial}{\partial \phi^p}$. In the
following all field indices (middle of the alphabet) will be raised
and lowered with the field metric and its inverse, and the covariant
derivative compatible with $\GG_{rs}$ will be denoted by
$\tilde{\nabla}$ to distinguish it from the space-time covariant derivative.

Equation \eqref{gen3c} is redundant as it is a consequence of \eqref{gen3a} and the $u$ derivative of Eq.~\eqref{gen3b}. Note that $2N+1$ {\it integration constants} are necessary to specify a solution of \eqref{gen3}: $2N$ for the Klein-Gordon equations \eqref{gen3a}, and 1 for the Einstein equation \eqref{gen3b}.  The number of integration constants will be important later when discussing non-gradient flows in the context of Hamilton-Jacobi (HJ) theory.

If the potential has an extremum at  $\phi=\phi_*$ where $V(\phi_*) =
- {d(d-1)}/{\ell^2}$, then  equations \eqref{gen3} have an  an
AdS$_{d+1}$ solution with constant scalars. This solution corresponds
to a  CFT. The  scale factor is $A(u)=-u/\ell$, where $\ell$ is the
curvature radius of AdS  and the boundary is at $u\rightarrow -\infty$.

\subsection{First order formalism}
\label{ss:1of}


For the following analysis, it will be useful to rewrite the equations
of motion in the first order formalism, first introduced in the
cosmological context for multiple scalar fields by Salopek and Bond
\cite{Salopek:1990jq}, and discussed  in great depth in the
holographic literature in e.g.~\cite{Skenderis:2006jq}).  An underlying assumption is that one
of the fields should (at least locally) have a monotonic evolution so
that it can traded for the $u$ coordinate, thus leading to $u$-independent first order equations. The starting point is the `fake superpotential' $W$, function of the field values $\phi^s$ only, and defined by
\eql{v2}{
W(\phi^s(u)):=-2(d-1)\dot A(u).
}
The existence of a superpotential is guaranteed for the single-scalar
case, piecewise in the regions where $\phi(u)$ is monotonic. 
As it has been argued \cite{Sonner:2007cp,Dorronsoro:2016pin}, and as we will discuss in more detail in
Section 3 this is true locally in field space also for the multi-field
case. Throughout this section we will simply assume that, for any
solution of the form (\ref{gen1}), a superpotential $W(\phi)$ satisfying
equation (\ref{v2}) exists.    We will return to this point in Section \ref{s:can} where we will critically asses this assumption. 

From action (\ref{action}), we define the field momentum densities by $\pi_r := \GG_{rs}\dot{\phi}^s$. They differ from the  canonical momenta by a factor of $-\sqrt{-g}$. Since the field-space metric is non-singular 
\eql{v1}{
\dot{\phi}^s = \GG^{rs}(\phi)\pi_r \equiv \pi^s(\phi).
}
We can rewrite equation  (\ref{gen3c}) using the definitions
(\ref{v2}) and (\ref{v1}), obtaining
\be
\pi^r\left(\pi_r - \de_r W\right) = 0\,. 
\ee
This implies that $\pi_r= \de_r W + \xi_r$, with $\pi^r\xi_r = 0$.  In
the special case $\xi_r=0$,  we are in the presence of a {\em gradient
  flow}. 

To characterise the non-gradient part $\xi_r$, we express  equations
\eqref{gen3a} and \eqref{gen3b} in terms of $W$ and $\pi_r$, 
\eql{v4}{
	\pi^q\WN_q \pi^p
		-{d\over 2(d-1)}W\pi^p
	-\GG^{ps}
	\p_pV
	=0,
}
\eql{v3}{
\ha \pi_r \pi^r-{d\over 4(d-1)}W^2-V=0.
}
Taking a
derivative of Eq.~\eqref{v3} and subtracting \eqref{v4} so as to eliminate  the potential $V$ leads to
\eql{v5}{
	\pi_p
	=
	\p_pW
	+
	{ 2(d-1)\over d~W} 
		\pi^s\FF_{sp}~, \qquad
                \FF_{sp}\equiv \WN_s \pi_p-\WN_p\pi_s.
}
It is thus clear that
\be \label{v5-i}
\xi_r = 	{ 2(d-1)\over d~W} 
		\pi^s\FF_{sr},
\ee
and  the flow is gradient  if
$\FF_{sp}=0$. 
In this section we only
restrict to gradient flows. The possibility of having non-gradient
flows will be considered in section \ref{ss:sym}. 

For gradient flows the independent equations become:
\begin{subequations}\label{eomW}
\begin{align}
	&\ha\GG^{rs}\p_rW\p_sW-{d\over 4(d-1)}W^2(\phi)=V(\phi)~,\label{SuperP}\\
	&\dot \phi^r(u)=\GG^{rs}(\phi)\p_sW(\phi)~,\label{floW}\\
	&\dot A(u)=-{1\over 2(d-1)}W(\phi)~.\label{floA}
\end{align}
\end{subequations}
Given a superpotential $W(\phi)$, integration of \eqref{floW} and \eqref{floA} introduces $N+1$ integration constants.  For a given potential $V(\phi)$, $W(\phi)$ itself is obtained by solving the partial {\it differential} equation (PDE) \eqref{SuperP},  referred to as the {\it superpotential equation}.
 A solution $W(\phi)$ to \eqref{SuperP} is specified by an {\it integration function} of $N-1$ variables.
 On the other hand, as we have seen  the total number of integration
 constants should be $2N+1$. Hence this formalism is highly redundant:
 the same flow is expected to arise from infinitely many
 superpotentials $W(\phi)$. It follows that it is enough to consider a
 subclass of solutions to the superpotential equation (\ref{SuperP})
 which contains  $N$ independent integration constants\footnote{In
   Hamilton-Jacobi theory this is referred to as a {\em complete
     integral}. We will make this connection in Section \ref{s:can}.}
 \cite{PapaCan}.  
%



\section{Holographic flows}
\label{s:hol}

We now focus our attention to holographic RG flows: these are
solutions which have an interpretation in gauge/gravity duality as
deformations away from conformality of a UV conformal fixed
point, which  corresponds to an extremum of $V$. Solutions  which are
everywhere regular connect the UV extremum to a second extremum of
the potential, interpreted as another conformal fixed point in the
IR.
For simplicity we
consider potentials which are strictly negative. This avoids
complications resulting from the bulk  curvature becoming small along
the flow, or from transitions to cosmological solutions. 

The holographic $\beta$-functions of the scalar couplings are given
by \cite{deboer,wenliang}
\be \label{curlyb}
\beta^r \equiv {\dot{\phi}^r \over \dot{A}} = -2(d-1)\GG^{rs}{\pi_s (\phi)\over W(\phi)}
\ee
which, the case of  gradient flows,  reduces to
\eql{h4}{
\b^r(\phi)
=-2(d-1)\GG^{rs}{\p_s W(\phi)\over W(\phi)}.
}
Before  proceeding, we  list two  important properties of the
superpotential $W(\phi)$. 
\begin{enumerate}
\item 
It follows from \eqref{SuperP} that $W$ is bounded from below by a positive function:
\eql{B}{
W(\phi)\geqslant B(\phi)>0,\qquad \text{where}\qquad B(\phi):=\sqrt{-{4(d-1)\over d}V(\phi)}.
}
where we have chosen $W>0$ without loss of generality (the
superpotential equation is invariant under $W\to -W$).  
This implies, through equation \eqref{floA}, that
$\dot A$ is always negative, so {\it $A(u)$ is monotonically
  decreasing with $u$}. 

\item The superpotential is  a monotonic function of the holographic coordinate $u$,
\eql{dWdu}{
{\dd W\over \dd u}=\dot \phi^r\p_rW=\GG^{rs}\p_rW\p_sW\geqslant0~.
}

\end{enumerate}

Property 2 implies that $W$ increases monotonically {\it along flows} and is stationary only when all
the $\b$ functions, defined in \eqref{h4}, vanish simultaneously,
implying $\dot{\phi^r}=0$.  
However this does not necessarily  mean
that the flow reaches a fixed point: for this one also needs
$\ddot\phi^r=0$ at the same point. By equation  (\ref{gen3a}), this can
only happen if $\p^r V=0$. 
 Therefore, true fixed points occur  
only when extrema of $W(\phi)$ are also extrema of $V(\phi)$. The case
of an extremum of $W(\phi)$ which is not an extremum of $V(\phi)$
corresponds to a {\em bounce}, i.e.~a regular point of the geometry where the flow is inverted in some of the directions.\footnote{In the single
  field case this was discussed in
\cite{multibranch}} Bounces will be
discussed in detail in subsection \ref{ss:bou}, and we now turn to the
analysis around extrema of $V(\phi)$. 
  
\subsection{Near-extremum analysis}
\label{ss:nex}
We start with a short review of the holographic dictionary around an
extremal point of the potential $V$ (assumed to be at the origin
without loss of generality). We assume $V$ has an analytic expansion
around the extremum, 
\eql{max2}{
	V(\phi)=-\frac{d(d-1)}{\ell^2}+ \sum_{r=1}^N
        {m^2_r\over 2}(\phi^r)^2 +\sum_{r,s,p,=1}^N \frac{g_{rsp}}{\ell^2}\phi^r\phi^s\phi^p+ \cO(\phi^4).
}
Note that we have chosen coordinates in field-space such that the mass matrix is
diagonal at the extremum \cite{Bourdier}, and we have
included terms up to cubic order in the scalar fields, controlled by arbitrary
 dimensionless coefficients $g_{rsp}$. 

Solutions of equations \eqref{gen3} with the potential \eqref{max2}
have the  asymptotic expansion 
\begin{subequations}\label{h7}
\begin{align}
&
A(u)=A_0-{u\over\ell}+\dots \label{h7a}
\\
&
\phi^r(u)=\phi^r_-e^{\D_r^- u/\ell}\left(1 +\dots\right)
+\phi^r_+e^{\D_r^+ u/\ell}\left(1 +\dots\right)~,\quad r=1,\dots,N, \label{h7b}
\end{align}
\end{subequations}
where
\eql{h8}{
\D_r^\pm={d\over 2}\pm\ha\sqrt{d^2+4m_r^2\ell^2}.
}
The parameters  $A_0$, $\phi_-^r$
and $\phi_+^r$ are   $2N+1$ integration constants, and hence fixing these
asymptotics completely determines the solution. One usually sets
$A_0=0$, corresponding to the boundary theory living on Minkowski
space with metric $\eta_{\mu\nu}$. The  parameters $\phi_-^r$
and $\phi_+^r$ are then related to the source $J^r$ and vacuum expectation values (VEVs) of the
corresponding  CFT operators by\footnote{We are working in the
  so-called {\em standard} dictionary. In the mass range $-d^2/4 < m^2 < -d^2/4+1$ there
  is an {\em alternative} dictionary, in which the roles of $\phi_+$
  and $\phi_-$ are interchanged.}:
\be\label{h9}
J^r = \ell^{-\Delta_r^-}\phi_-^r, \qquad \le<\cO_r\ri>_J=(2\D^+_r-d) \ell^{-\Delta^+_r}\phi^r_+ \, ,  
\ee
and $\Delta^-_r$ and $\Delta^+_r$ are interpreted  as the conformal
dimension of the source $J^r$ and the operator $\cO_r$,
respectively. 

While $\Delta_r^+ >0$, $\Delta^-_r$ can have either sign
depending on whether the operator $\cO_r$ is relevant or
irrelevant. For an extremum with $M$ negative  and $N-M$
positive mass eigenvalues   $m_r^2$,  it is convenient to split the $N$
directions in field space into two sets, $\hat{r} = 1\ldots M$ and
$\check{r} = M+1\ldots N$,  and to introduce the notation\footnote{The symbols are chosen so that $\hat{r}$
labels direction along which $\phi=0$ is a maximum, whereas
$\check{r}$ labels those directions along which the extremal point is a
minimum.}:
\be\label{Deltas}
 \left\{\begin{array}{llll}
 	\D_{\check{r}}^- \leqslant 0~,
	 & \quad m_{\check{r}}^2 \geqslant 0~,
	 &  \quad  \check{r} =
     M+1,\ldots, N &  \quad \text{irrelevant operator.}
     \\
     & & & 
     \\
  0 \leqslant \D_{\hat{r}}^- < {d\over 2} ~,
  & \quad -{d^2 \over 4\ell^2} <  m^2_{\hat{r}} \leqslant 0 ~,
    &  \quad  \hat{r} =
     1,\ldots ,M & \quad \text{relevant operator.}\end{array}
\right. 
\ee
The lower bound on negative values of $m^2$ is the Breitenholer-Freedman bound and it is
required for perturbative stability of the solution.

The expansions (\ref{h7a}-\ref{h7b}) must hold as $\phi \to 0$.  Depending on the signs of the
$\Delta^-_r$, this  corresponds to either
$u\to+\infty$ or $u\to -\infty$. 
The  allowed 
combinations of integration constants are
\begin{itemize}
\item[{\bf UV:}]
\be \label{UV1}
 u \to -\infty, \; \exp A(u) \to +\infty ,  \qquad \phi_-^{\hat{r}}\neq  0, \; \phi_-^{\check{r}} = 0 ,\;
\phi_+^r \;\; \text{arbitrary}; 
\ee
From the field theory perspective,  we can only turn on the sources $\phi_-^{\hat{r}}$
corresponding to {\em
  relevant} operators, for which $\Delta^-_{\hat{r}} >0$, and we have
to set to zero those corresponding to {\em irrelevant} operators
($\Delta^-_{\check{r}} < 0$).  Also, we are free to turn on any
combination of vevs $\phi_+^r$.   
\item[{\bf IR:}]
\be\label{IR1}
u \to +\infty, \; \exp A(u) \to 0,  \qquad \phi_-^{\check{r}}\neq 0, \;   \phi_-^{\hat{r}} =  \phi_+^r = 0 
\ee
In this case only the sources $\phi_-^{\check{r}}$ for the {\em
  irrelevant} operators can be non-zero, because the IR fixed point is
stable under deformations by such operators. However turning on any
other source or vev would make the flow miss the IR fixed point. 
\end{itemize}

In the following we study flows around the extrema from the point
of view of the superpotential equation rather than the equations of motion. In the single-field case, close to an
extremum of $V$  the
superpotential has a universal analytic term (which can be of two
different kinds), plus a sub-leading
non-analytic piece which contains the single integration constant to the
superpotential equation \cite{wenliang,multibranch}. 
As we will see in the next two subsection,
this structure persists in  the multi-field case, with the difference
that there are many more branches of analytic solutions, and a larger
class of non-analytic deformations parametrised by an integration function.

\subsubsection{Analytic part of  the superpotential}

  We start by looking for {\it analytic} solutions $W_0(\phi)$ for equation
\eqref{SuperP} around the origin, with the potential $V(\phi)$ in equation
\eqref{max2} such that both have extrema at $\phi^r=0$, following \cite{Bourdier}.
We use Riemann normal coordinates on the scalar manifold ${\MM}_{\phi}$
\be
	\GG_{rs}(\phi^s)
	=\d_{rs}
	-{1\over 3}\RR_{rpsq}\phi^p\phi^q
	-{1\over 6}\le(\WN_m\RR_{rpsq}\ri)\phi^p\phi^q\phi^m
	+\cO(\phi)^4\, ,
	\label{RNor}
\ee
  where $\RR_{rpsq}$ is the Riemann curvature tensor associated with
  $\GG_{rs}$.

Expanding both sides of equation \eqref{SuperP} in powers of $\phi^r$
around $\phi^r=0$,
it is easy to show that there are $2^N$  analytic  solutions of equation \eqref{SuperP},
parametrised by a string $\sigma = (\sigma_1, \ldots
,\sigma_N)$ where $\sigma_i = \pm$,   
\be\label{max4b}
W_0^{\sigma}(\phi)
	={2(d-1)\over \ell}
	+{1\over 2\ell}\sum_{r=1}^N\Delta_r^{\sigma_r}\le(\phi^r\ri)^2
	+\sum_{p,q,r=1}^N
	{ g_{pqr}\phi^p\phi^q\phi^r\over\ell  (\Delta_p^{\sigma_p}+\Delta_q^{\sigma_q}+\Delta_r^{\sigma_r} - d)}
	+\cO\le(\phi^4\ri),
\ee
These  $2^N$ solutions are the generalisation to the multi-field case
of the two branches $W^+$, $W^-$ for a  single field
\cite{multibranch}. When the $\Delta's$ in the denominator of the
cubic term sum up to $d$, logarithms will appear in the solution. 

\begin{figure}[t]
\centering
\includegraphics[width=0.45\textwidth]{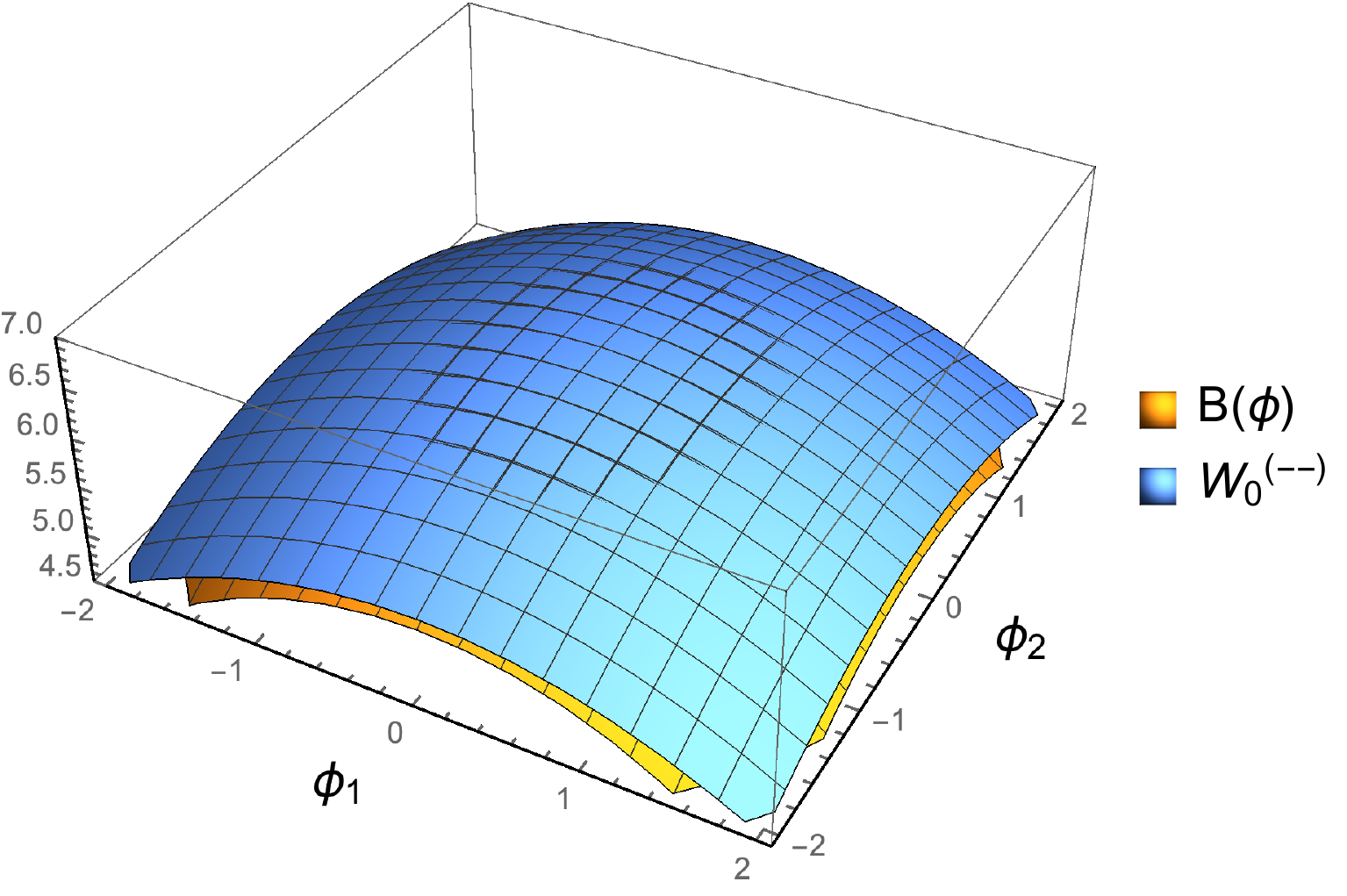}
\includegraphics[width=0.45\textwidth]{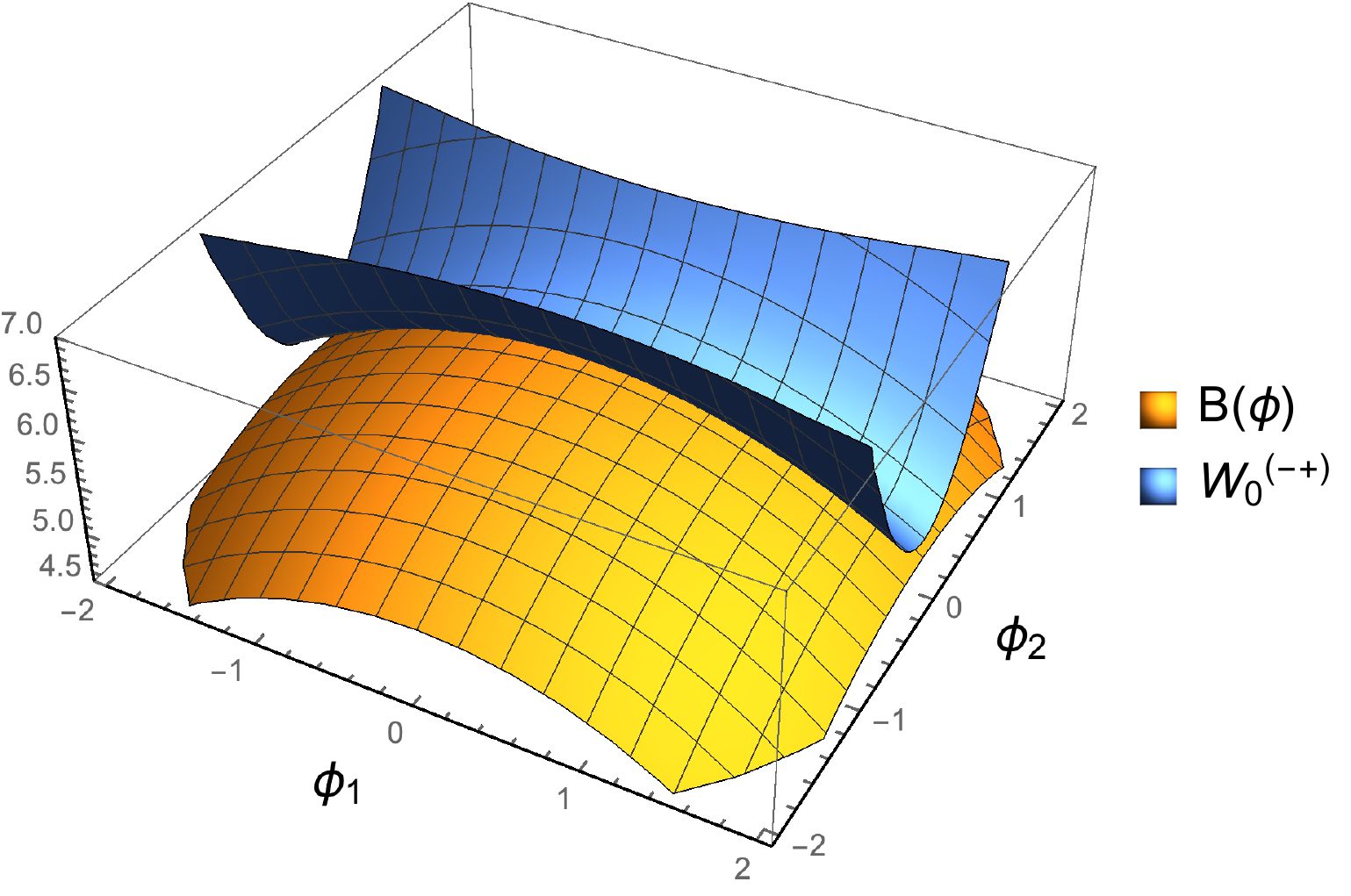}
\includegraphics[width=0.45\textwidth]{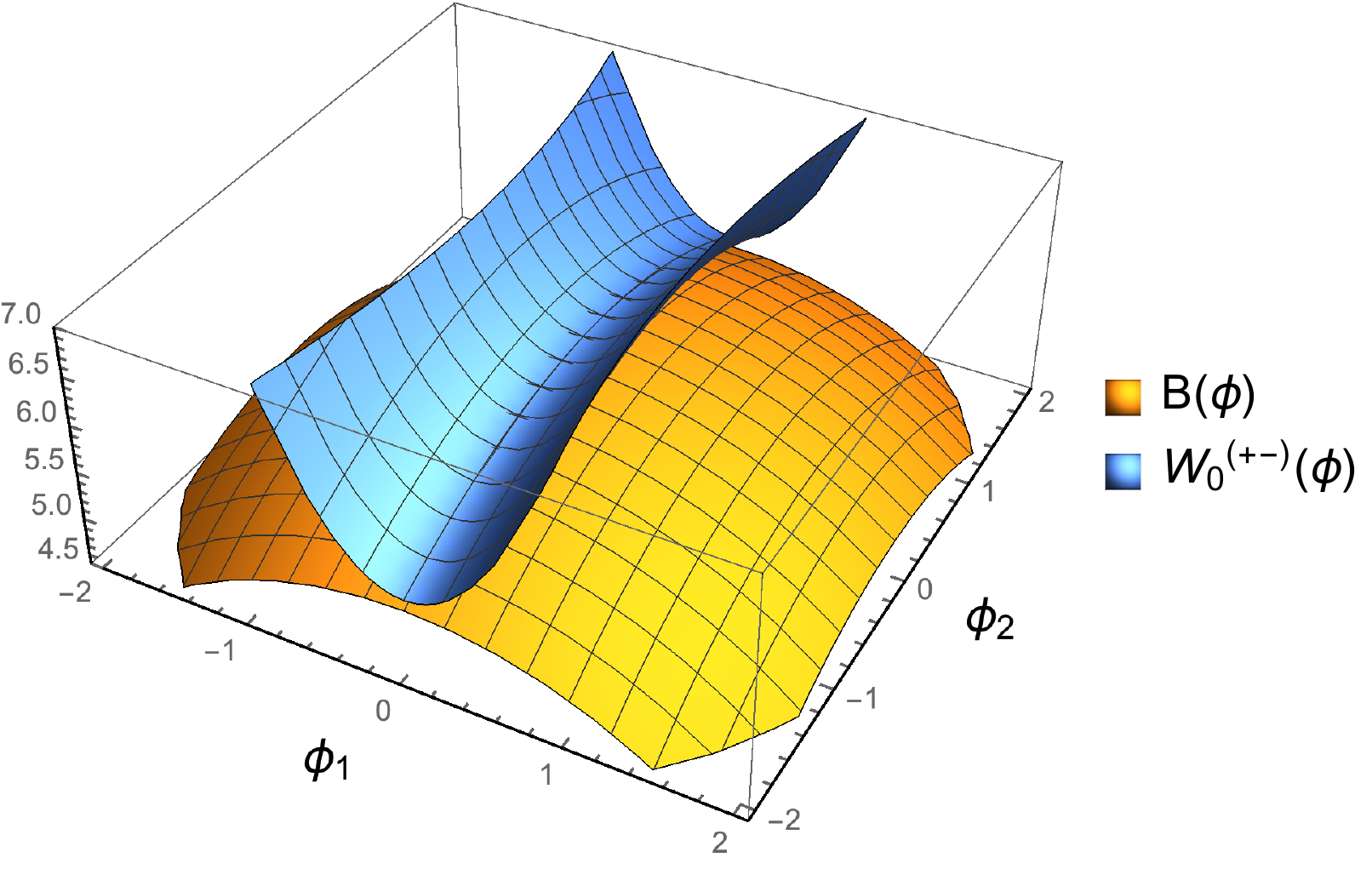}
\includegraphics[width=0.45\textwidth]{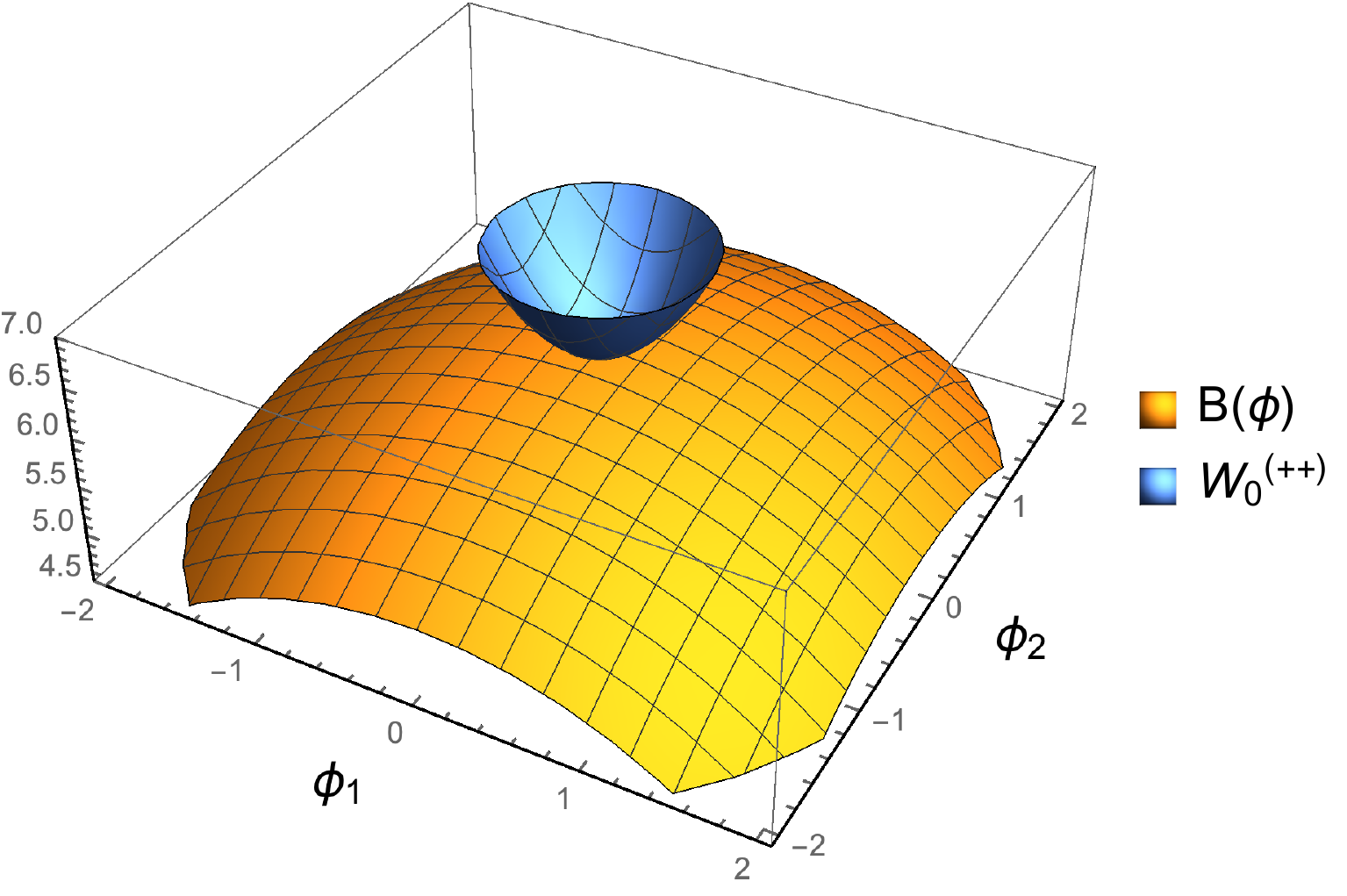}
\caption{This figure shows the four leading solutions of the
  superpotential equation close to a
  minimum of the potential,   of the form
  (\ref{max4b}),  in the two-field case. The yellow surface is the critical
curve $B(\phi)$. 
} 
\label{f:minW}
\end{figure}

All  solutions $W^\sigma_0$ have an extremum, which may be a maximum, minimum, or a saddle point, at $\phi=0$. However this extremum of $W^\sigma_0$ does {\it not} necessarily have the same
signature (of the Hessian matrix) as the extremum of
$V$. The latter is characterised by the signature of the mass matrix, 
 which determines whether   the corresponding operators are relevant
 ($\phi^{\hat{r}}$)  
 or irrelevant ($\phi^{\check{r}}$).  The former, $W$, is determined
 by the signs of the $\Delta^\pm_r$. These are not necessarily
 the same as  the signs of $m^2_r$ since $\Delta^+_r >0$ for any
 mass, see equation (\ref{Deltas}). This is illustrated in figure
 \ref{f:minW}: the {\em same} minimum of $V$ admits four different solutions $W_0^\sigma$  with different
 signature of the Hessian matrix, which in our notation are $W^{(++)}_0,
 W^{(+-)}_0, W^{(-+)}_0, W^{(+-)}_0$.    

Close to $\phi=0$ the flow equations arising from one of the
superpotentials $W^\sigma_0$  become linear and decoupled, 
\be\label{flow1}
\dot{\phi}^r \approx {\D_r^{\sigma_r} \over \ell} \phi^r, \qquad \dot{A} \approx -{1\over \ell}. 
\ee 
To use the language of dynamical system,  we  may
have both {\em repulsive} directions  ($\Delta^{\sigma_r}_r >0$)  and
{\em attractive} directions  ($\Delta^{\sigma_r}_r < 0$). A repulsive
direction can correspond  either to a relevant operator
($r=\hat{r}$, and $\sigma_{\hat{r}} = \pm$) or to an irrelevant operator 
($r=\check{r}$ with the choice $\sigma_{\check{r}} = +$). An attractive
direction always correspond to an irrelevant operator with the choice
$\sigma_{\check{r}} = -$. 

\begin{figure}[t]
\centering
\includegraphics[width=0.45\textwidth]{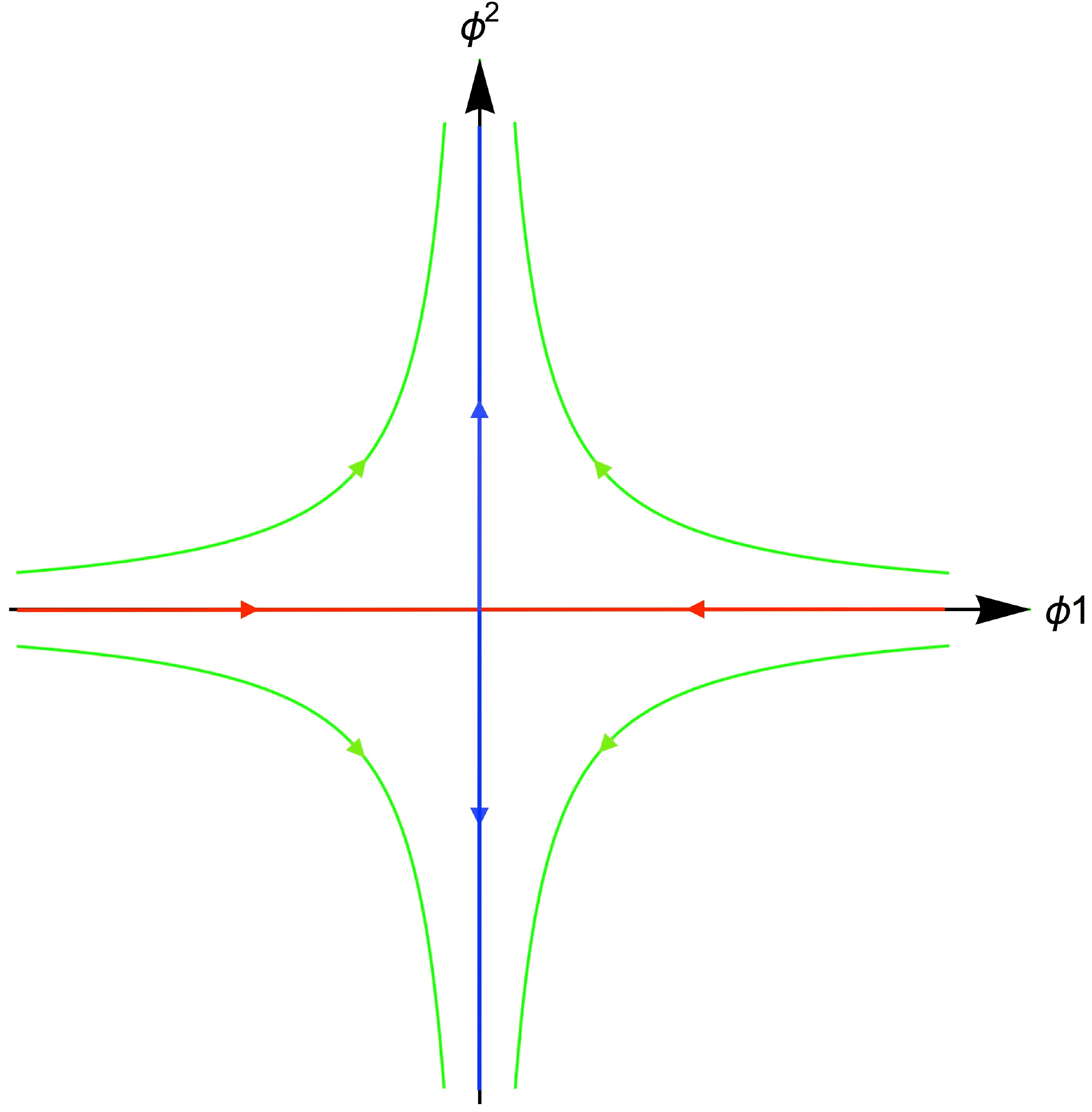}
~~
\includegraphics[width=0.45\textwidth]{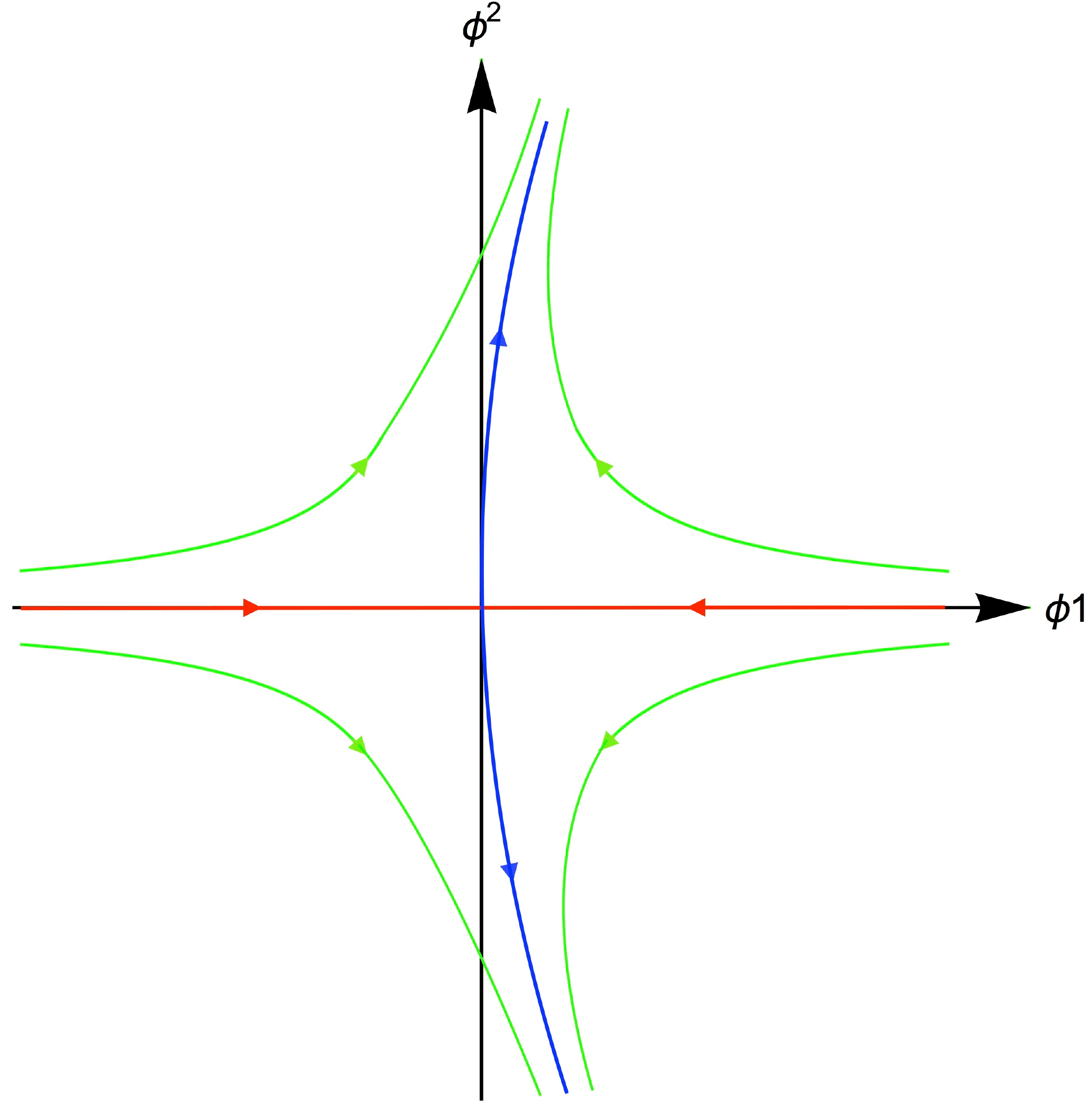}\\
(a) \hspace{0.45\textwidth} (b)
\caption{The left figure represents flows associated with a $W^{(+-)}_0$ solution around a local minimum of $V(\phi)$ as depicted in figure \protect\ref{f:minW} {\it or} the flows from a  $W^{(--)}_0$ solution of a saddle of $V(\phi)$ with $m_1^2>0$ and $m^2_2<0$. The right figure represents a possible effect of the non-linear terms which may or not include non-analytic terms.
} 
\label{f:saddle}
\end{figure}

When $W^\sigma_0$ has   both repulsive and
attractive directions, generically the fixed point will not be reached, 
since  generic initial conditions  will violate both conditions
(\ref{UV1}) and (\ref{IR1}): the flow may approach the fixed point but
miss it and leave along another direction. This situation is
represented in the two-field case in figure \ref{f:saddle} (a), which
represents the  flow diagram of a solution of the type $W_0^{(-+)}$
with $\Delta^-_1 <0$.  
Nevertheless, there are special (fine-tuned) initial
conditions  (namely $\phi_-^1=0$ or $\phi_+^2 = 0$)   which satisfy either (\ref{UV1}) or (\ref{IR1}),
and such that  the flow will reach $\phi=0$ in the UV (blue line) or the IR (red
line), respectively.    

Notice that the same  solution (\ref{max4b}) with a given $\sigma$  can describe both a UV and an IR
fixed point, depending on the choice of initial conditions for the
first order flow (\ref{flow1}) . This is very different from  the single field case, when each branch around an
extremum can be unambiguously assigned to the UV or the IR, independently
of the initial conditions for the first order  flow \cite{multibranch}.


\subsubsection{Deformations of the analytic solution}
As mentioned earlier,  a solution $W(\phi)$ of equation (\ref{SuperP})
is specified by an integration function of $N-1$ variables. The
$2^N$ analytic solutions \eqref{max4b} on the contrary, do not depend on any continuous parameter. The reason is that, on any of the $2^N$
branches, 
the leading analytic behaviour close to a fixed point is universal, and the
difference between solutions arises through sub-leading, non-analytic
terms \cite{PapaMulti,wenliang}. Therefore we now consider subleading deformations\footnote{Throughout this section  we have assumed $m^2_r\neq0$. The case when
one of the masses vanishes, corresponding to $\Delta_-=0, \Delta_+=d$,
(i.e.  the  operator is
marginal) can be treated along the same lines, as it was done for the
single field case in \cite{multibranch}.} of the
$2^N$ solutions $W_0^\sigma$ of \eqref{max4b}, 
\eql{max5}{
W(\phi)=W_0^\sigma(\phi)+\d W(\phi). 
}
The linearised  equation for   $\d
W$ following from  (\ref{SuperP}) is
\eql{max6}{
\GG^{rs}\p_rW_0^\s\p_s \d W={d\over 2(d-1)}W_0^\s \d W
}
which, to leading order in powers of the the scalar fields, becomes
\eql{der1}{
\sum_{r=1}^N\D_r^{\s_r}\phi^r\p_r \d W = d \d W.
}
The general solution is a
linear superposition of separable solutions 
\eql{der2}{
\d W_S=f_1(\phi^1)f_2(\phi^2)\dots f_N(\phi^N).
}
From \eqref{der1}, the $f_s$ are given by
\eql{der4}{
f_s=\abs{\phi^s}^{\kappa_s/\D_s^{\s_s}},\qquad \sum_{s=1}^N \kappa_s=d.
}
where $\kappa_s$ are $N$ constants. 
Therefore, to leading
order around $\phi^r=0$, the general solution to \eqref{max6} is a superposition of all the  special solutions \eqref{der4}, namely
\eql{max7}{
\d W=  \frac{1}{\ell} \int d \kappa_1\dots \int d\kappa_N\prod_{r=1}^N  \le(\phi^r\ri)^{\kappa_r/\Delta_r^{\s_r}}~K(\kappa_1,..,\kappa_N)~
\d\le(\sum_{s=1}^N \kappa_s-d\ri).
}
where $K(\kappa_i)$ is a function (or more generally a distribution)
which is largely arbitrary. This must be subject to some constraints,
for example it needs to have support on those $\kappa_r$ which have the
same sign of the corresponding $\Delta_r$, orthewise $\delta W$ will not
vanish close to $\phi^r=0$. There will also be bounds on $K$ at
infinity from the convergence of the integral, whose detailed
analysis is beside the point here.     What is important is that equation (\ref{max7}) gives  a huge
multiplicity of deformations.

An significant qualitative difference with
respect to the single field case is the following. In the case of only
one scalar,  solutions of the type $W^+$ and infra-red
solutions of the type $W^-$ (reaching a minimum of $V$) are isolated
and do not allow continuous deformations. As a consequence, imposing
that $W(\phi)$ reaches a minimum of the potential fixes the
superpotential completely, and once this choice is made all flows will
automatically reach the IR fixed point. 
In the multi-field case instead, solutions with  any number of
$\Delta_r^+$ or negative $\Delta_r^-$ still admit deformations. 

There is one case when the deformation is 
forbidden.  When $\phi=0$ is a {\em minimun} of $V$, all $\Delta^-_r
<0$. The solution $W^{(-,-\ldots -)}_0$ in (\ref{max4b}) corresponding
to the choice $\sigma_r = -$ for all components has only attractive directions, with two consequences:
1) a generic  flow arising from this superpotential reaches the IR fixed
point; 2) the deformation is identically zero because the
$\delta$-function cannot be saturated while having all $\kappa_r<0$  at the
same time, which is required in order that only positive powers of the
fields arise. These two facts have very important consequences, as we
will discuss at the end of this section in subsection \ref{sss:reg}.

The deformation $\delta W$ can turn on a flow velocity even in those directions
which, to leading order, where attractive, by giving vevs to the
corresponding operators by mixing. We illustrate this with a simple
two-field example. 

\paragraph{Two fields example.} Consider a saddle point with one
irrelevant  ($m^2_1>0$) and one relevant  ($m^2_2<0$) direction, and
choose the analytic solution with $\sigma = (--)$: 
\be\label{2f1}
W_0^{(--)}(\phi)
	={2(d-1)\over \ell} + 
	{\Delta_1^{-}\le(\phi^1\ri)^2\over 2\ell} +
        {\Delta_2^{-}\le(\phi^2\ri)^2\over 2\ell}+ O(\phi^3),
	\quad
	\D_1^-<0~,
	\quad
	0<\D_2^-<{d\over2}
	~.
\ee
We will neglect cubic term for now (we will comment on their effect later). 
The flow diagram is again represented by figure \ref{f:saddle} (a). 
Close to the origin, to this order, the flow equations \eqref{floW}
are 
\be\label{2f2}
\dot \phi^1={\D_1^-\over\ell}\phi^1+\dots
~,
\qquad
\dot \phi^2={\D_2^-\over\ell}\phi^2+\dots
~.
\ee
and their general solution is
\be\label{2f3}
\phi^1=\phi^1_-e^{\D_1^-u/ \ell}+\dots,
\qquad
\phi^2=\phi^2_-e^{\D_2^-u/ \ell}+\dots
\ee
We can take  the limit of $u \to \pm \infty$ only if either
$\phi^1_-=0$ (UV fixed point) or $\phi^2_-=0$ (IR fixed point). In the
single field case, it would be impossible to turn on a  non-analytic
deformation corresponding to a vev along
the attractive direction  $\phi^1$.  
Now we will show that, instead,  a non-vanishing $\phi^1_+$  can be can be generated by $\phi^2_-$ through the following non-analytic deformation:
\be\label{2f4}
	\d W
	=
	C_1 \phi^1\le(\phi^2\ri)^{\D_1^+/\D_2^-}, 
\ee
which corresponds to a separable deformation of the kind (\ref{der2})
with  $\kappa_1 = \Delta_1^-$, $\kappa_2 = \Delta_1^+$. Notice that
this is allowed since by definition $\Delta_1^-+ \Delta_1^+=d$. 

Adding the sub-leading term (\ref{2f4}) to (\ref{2f1})  changes the flow equations to:  
\bea
&&\dot \phi^1
=
	\D_1^-\phi^1
 + C_1\le(\phi^2\ri)^{\D_1^+/\D_2^-}
 \label{2f5} \\
&&
\dot \phi^2
=	{\D_2^-}\phi^2 
\label{2f6}
\eea
Integration of \eqref{2f5} and \eqref{2f6} lead to the following expansions where only the first non-zero contribution from $W_0$ and from $\d W$ is presented for each field:
\begin{align}
\label{2f7}
&	\phi^1(u)
	= {C_1\over 2\D_1^+-d} \le(\phi^2_-\ri)^{\D_1^+/\D_2^-}
	e^{\D_1^+u/ \ell}
	+
	\dots
\\
&	\label{2f8}
	\phi^2(u)
	=
	\phi^2_-e^{\D_2^-u/ \ell}
	+\dots
\end{align}
We can see from equation (\ref{2f7}) that the deformation (\ref{2f4}) has the
effect of generating a vev-type term  for $\phi^1$ from a source-type
term for $\phi^2$, 
\be
\phi_+^1 = {C_1\over 2\D_1^+-d} \le(\phi^2_-\ri)^{\D_1^+/\D_2^-}.
\ee
Even if we start on the $\phi^2$  axis, $\phi^1$
will start to flow, as shown in figure (\ref{f:saddle}(b)). Notice
that, although it is possible to generate the same effect
``perturbatively'' using the cubic terms in the analytic solution, one
will not obtain the correct scaling for a vev-type sub-leading
term.

 Notice that the same  flow (\ref{2f7}-\ref{2f8}) can also be obtained
 starting from a solution of the kind $W_0^{(+-)}$ {\em without}
 deformation. This is an example of the redundancy of the
 superpotential description, and of the fact that the same flow can be
 obtained from different superpotentials. 
\\

We now return to the general $N$-field case.
Given this redundancy, it is desirable to find a ``minimal'' set of
deformations which encodes all possible flows that reach an extremum of
the potential, and depends on a finite
number of parameters. Such a minimal set is given by
\eql{max9}{
\d W=\frac{1}{\ell}\sum_{r=1}^NC_r\le(\phi^r\ri)^{d/\Delta_r^{\sigma_r}}.
}

The expression \eqref{max9} is essentially the  sum of $N$ single-field
deformations. In this case we have to impose the restriction
\be
0< \Delta_r^{\sigma_r} < d/2,
\ee
because we want all terms in equation (\ref{max9}) to have powers
which are both positive and sub-leading with respect to the  leading quadratic
terms in $W_0$. Since  by equation  (\ref{h8}) $\Delta^+_r > d/2$, we
conclude that the deformation of the form (\ref{max9}) have to satisfy:
\eql{max10}{ \text{either} \quad
\le\{\D_r^{\sigma_r}=\D_r^- \quad\nd\quad
\Delta_r^->0\ri\}\qquad\text{or}\quad C_r=0, 
}
i.e, as in the single field case,  it can only appear on for {\em
  relevant} operators for which the leading term in $W_0^\sigma$
corresponds to turning on a {\em  source} $\phi_-$.    
 The constants $C_r$, as in the single field case, determine
the sub-leading ($vev$) terms in these directions, 
\be \label{Cvevs}
\phi_+^r = { C_r  d \over \Delta^-_r\Delta^+_r} \le(\phi_-^r
\ri)^{\Delta_r^+/\Delta_r^-}. 
\ee

It is easy to check that, for a generic UV extremum with $M$ relevant
and $N-M$ irrelevant directions, the deformations in
the class (\ref{max9})  are enough to have 
as many  integration  constants as there are  boundary conditions satisfying the
constraint (\ref{UV1}): there are $M$ $C_{\hat{s}}$ in $\delta W$,
plus $N$ initial conditions for the flow equations. These correspond to
 the $M$ source terms $\phi^{\hat{r}}_-$ and $N$ vev terms
$\phi^{\hat{r}}_+$ which are needed to fix the asymptotics in the UV,
see equation  (\ref{UV1}). The remaining $N-M$ integration constants
(which should in principle be there) are necessarily set to zero if the
flow has to reach the UV fixed point. Thus the deformations of the
type (\ref{max9}),  constrained by (\ref{max10}),  are enough to
obtain all   (non-generic) solutions which reach an extremum. 

If we start from an extremum which is a local {\em
  maximum} of the potential $V$, all the $\Delta_r^-$ are positive and
the deformation (\ref{max9})  contains $N$ arbitrary integration
constants $C_r$, which is the number needed to describe a generic
solution. This is what we will refer to in the next section as a {\em complete
  integral}, to use the terminology of Hamilton-Jacobi
theory. Therefore solutions of this form which arrive at UV maxima are
generic\footnote{This does not mean that {\em all} solutions will
  arrive at a given maximum, e.g. if there are multiple local
  maxima. However one does not need to tune the integration constants
  to arrive there, unlike the case for generic extrema.}. 
In the case of a {\em  minimum}, all $m^2_r$ are positive
and all the $C_r$ in (\ref{max9}) are constrained to vanish. 

In table (\ref{car10}) we summarise the allowed deformations of the special
type (\ref{max9}) and  the interpretation of the corresponding
operator in the holographic language. 

\begin{table}[h]\label{car10}
\centering
\begin{tabular}{|c|c|c|c|c|c|c|}
   \hline
    $\Delta $ & $m^2$ & type of fixed point & type of operator &
   source & vev & Deformation \\
   \hline
   $+$ &  $>$ & UV & Irrelevant & 0 & $\neq 0$  & No \\
   \hline
   $+$ & $<$ &UV & Relevant & 0 & $\neq 0$ & No \\
   \hline
   $-$ & $>$ &IR & Irrelevant & $\neq 0$ & 0  & No \\
   \hline
   $-$ & $<$ & UV & Relevant & $\neq 0$ & $\neq0$ & Yes \\
   \hline
\end{tabular}
\caption{A classification of the deformation  of the type (\ref{max9})  in terms of the signs of the squared masses and the choice of $\D^+_r$ or $\D^-_r$. The nature of the fixed point when reached along $\phi^r$, i.e., where all the $\phi^s$ with $s\neq r$ are set to zero, is indicated. When all the masses have the same sign and the dimensions are of the same kind, we can unambiguously classify the fixed points, otherwise a single extremum may belong to different categories above.}
\end{table}

To summarise, around  an extremum of $V$ with  $M$ negative and $N-M$ positive
mass eigenvalues, the set of superpotentials
\bea
W^\sigma = && {2(d-1)\over \ell} 
	+{1\over 2\ell}
	\le(
		\sum_{r=1}^N \Delta^{\sigma_r}_r\le(\phi^r\ri)^2
		+\cO\le(\phi\ri)^3
	\ri) \nonumber \\ 
&& +\sum_{\hat{r}=1...M\, \text{and}\, 
                    \sigma_{\hat{r}} = -}
C_{\hat{r}}\le(\phi^{\hat{r}}\ri)^{d/\Delta_{\hat{r}}^-} 
\Big(1	+	\cO(\phi^3) \Big) \label{special}
\eea
encodes all flows arriving to or departing from the extremum. It
is important to remark that, when considering solutions which connect 
different extrema, generically one can choose the special form
(\ref{special}) only close to one of them. For the others, the
deformation will have a more general form of the type (\ref{max7}).

If we include the sub-leading  cubic term in $W_0^\sigma$ from equation  (\ref{max4b}) in solving equation
(\ref{max6}), the corresponding solution reads
\eql{2dW}{
\d W(\phi)=\sum_{\Delta = \Delta_{\hat{r}}^-}C_s\le(\phi^s\ri)^{d/\Delta_s-1}
			\le[
				\phi^s
				+\sum_{p,q=1}^N  {3~d~ g_{spq}~ \phi^p~\phi^q\over \D_s \le( \Delta_s+\Delta_p+\Delta_q-d \ri)\le(\D_s-\D_p-\D_q\ri)}
			+
			\cO(\phi)^3
			\ri]
~.
}
The denominators of the cubic term
can vanish for certain combinations of dimensions, in which case the
corresponding terms in the expansion are replaced by terms containing
logarithmic contributions. 
There are further corrections to this expression, organised in a
double series expansion in $\phi^r$ {\em and} $C_r$. The general form
in the single field case can be found in \cite{Bourdier}.

\subsubsection{Lifting the arbitrariness in $W$: IR regularity} \label{sss:reg}

As we have seen, the solution $W^{(-,-\ldots -)}$ at a minimum of  $V$
does not admit continuous deformations. This fact has very far reaching consequences. In fact, as we will argue below,  $W^{(-,-\ldots -)}$ is the
{\em only} one among all the $W^{\sigma}$ which fulfills an important
regularity condition.  If this condition is imposed, the uniqueness of $W^{(-,-\ldots -)}$  lifts the
redundancy in the superpotential description and completely fixes the
solution $W$  everywhere in field space. Thus, the theory has only one
physical vacuum (or at most a finite number, if there is more than one
local minimum of $V$) for any value of the UV initial conditions for
the flow (which do not enter in the superpotential).

We now discuss the regularity condition. In holography, not all bulk solutions are
allowed, but only those which satisfy certain  conditions in
the IR. If one requires strict regularity (finiteness of the curvature invariants), then in the class of
  solutions (\ref{gen1}), the only possibility is that the flow
  reaches an IR asymptotically AdS fixed point. However, one
  does not want to 
  check regularity individually for each flow. This is where the
  superpotential formulation is useful: it gives a way of imposing
  regularity for whole classes of flows at the same time. With this in
  mind, the most economic regularity requirement is that {\em all}
  flows around a minimum of $V$ reach the minimum as an IR  fixed
  point.  This is the case if, around the minimum, $W$ is chosen to be
  $W^{(-,-\ldots -)}$.    This choice is reasonable, because it leaves the possibility
  of slightly deforming the initial conditions in the UV (i.e. the
  values of the UV sources) without
  spoiling the IR behavior. This is crucial from the dual QFT point of view: for example, every time
  we compute a correlation function  we perturb the sources, and we do
  not want to worry about the fact that a small deformation may render
  the theory inconsistent. Of course, a big change in the source may
  still lead the flow elsewhere, therefore ultimately the range of
  allowed values may be restricted. But this is not unusual even in
  perturbative field theories, where it is known that certain theories
   make sense only in certain {\em continuous} ranges of couplings
   (e.g.~for  $\lambda \phi^4$ theory we must restrict to $\lambda
   >0$). The fact that not all ranges of couplings lead to
   IR-regular solutions was also observed in holographic RG flows in
   \cite{multibranch}: there, we saw an example where for negative UV
   source an IR fixed point may be reached, whereas for positive UV
   source no regular solution exists.

To summarize, imposing that the superpotential around a minimum of $V$
has the form $W^{(-,-,\ldots,-)}$ completely fixes the full $W$ while
at the same time not restricting the values of the UV sources. 

Of course, if we impose this condition, there is no guarantee that the
solution in the UV will look like one with  the special  class of
deformation, (\ref{special}). In general, the subleading non-analytic
part will have the form (\ref{max7}), with some fixed  function $K(\kappa_r)$.

\subsection{Bounces}
\label{ss:bou}

We now study the geometry close to an extremum of $W$ which is {\it not} an
extremum of $V$. At these points, the flow inverts its direction (it
``bounces''). 
In the single-field case this behavior was analysed in detail in
\cite{multibranch} where it was shown to lead to a breakdown of the
first-order formalism and to multi-branched superpotentials.  In the
multi-field scenario, there is a much  richer a variety of bounces.  We befin with a brief review of the single field case.

\paragraph{Review of single-field bounces}

Here we set $N=1=\GG_{11}$ to one as can always be done for a single-field. The superpotential equation is
\eql{BN1}{
V=\ha W'^2-{d\over 4(d-1)}W^2
}
where $W'=dW/d\phi$.
Suppose  $W'$ vanishes at a point  $\phi=\phi_B$ where the potential
has a regular expansion and non-vanishing first derivative. 
Then, expanding  the superpotential equation (\ref{BN1}) around
$\phi_B$ one finds 
 two branches of $W(\phi)$ that meet at $\phi_B$. For concreteness consider $V'(\phi_B)>0$. 
There are two solutions connecting to $\phi_B$, 
\bea
 W_{\uparrow}(\phi)\simeq  W_B+\frac{2}{3} \sqrt{2V'(\phi_B)}(\phi-\phi_B)^{3/2}, \nonumber \\
&& \quad \phi > \phi_B, \label{BN6}\\
 W_{\downarrow}(\phi)\simeq  W_B-\frac{2}{3} \sqrt{2V'(\phi_B)}(\phi-\phi_B)^{3/2},  \nonumber
\eea
where
\eql{BN5}{
W_B\equiv W(\phi_B)=\sqrt{-{4(d-1)\over d}V(\phi_B)} \equiv B(\phi_B).
}
Equation (\ref{BN5}) means that bounces occur when the flow reaches
the critical curve which bounds the forbidden region defined in equation
(\ref{B}).  	
The two branches can be merged into a fully regular solution for the
metric and the scalar: integrating the flow equations  we obtain
\begin{align}
	\phi(u)=\phi_B+{V'(\phi_B)\over 2}(u-u_B)^2+\cO(u-u_B)^3=
		\begin{cases}
			\phi_\uparrow(u)\text{ for } u>u_B,\\
			\phi_\downarrow(u)\text{ for } u<u_B, 
		\end{cases}
	\label{phiB}
\end{align}
\be \label{Ab}
A(u)=A_B-\sqrt{V(\phi_B) \over d(d-1)}(u-u_B)
+\cO(u-u_B)^4,
\ee
It is clear from \eqref{phiB} and \eqref{Ab} that at the point
$\phi_B$ both the scalar field and the metric are regular. The bounce corresponds to a point where the first  (but not
the second) derivative $\dot\phi(u)$ vanishes at some point $u=u_B$, and the superpotential
becomes double-valued because $\phi$ ceases to be a good
coordinate. The two branches correspond to $u > u_B$ and $u<u_B$, and
$\phi< \phi_B$ on both branches.

\paragraph{Multi-field bounces}

When many flows of a family have a field component
which is reversed along its flow, the velocity field is multi-valued.
 This means that a superpotential
required to describe such flows has different branches. We will call a
\emph{bounce}  a point (or a set of points) in field
space around which  a solution
$W(\phi)$ to the superpotential equation \eqref{SuperP} has more than one
branch, i.e. some of the scalars reverse their direction along the
flow.  When $n$ scalars reverse direction simultaneously we call this
an {\it order $n$ bounce}. The special case of $n=N$ will be called a
{\it complete bounce} to  distinguish it from the cases where $n<N$,
which we will call {\it partial bounces}.

\subsubsection{Complete bounces}
\label{sss:cba}

Complete bounces are defined as loci $\phi=\{\phi_B^r\}$ where a solution $W(\phi)$ of the superpotential equation \eqref{SuperP} reaches its lower bound, 
\eql{b0}{
	W_B \equiv W(\phi_B)=\sqrt{-{4(d-1)\over d}V(\phi_B)}\equiv B(\phi_B)
~.
}
These may be isolated  points, or  may form a co-dimension $p$
sub-manifold, $\Sigma_{N-p}$, of the scalar manifold $\MM_\phi$. In
the latter case  $\Sigma_{N-p}$ lies on an equipotential surface of
$V(\phi)$, since by the superpotential equation \eqref{SuperP} all
derivatives of $W$ must vanish at all points on $\Sigma_{N-p}$. 
Hence $W$, and thus $V$ from (\ref{b0}) is constant.



Because an equipotential is a co-dimension-one sub-manifold of
$\MM_\phi$, the  simplest example of a bounce is obtained by giving
the superpotential equation  initial condition \eqref{b0} on an entire
connected piece of an equipotential\footnote{Typically, an
  equipotential is the union of disjoint hyper-surfaces and to
  consider bounces in the sequence we will only consider connected
  equipotentials.}, which we will denote $\Sigma_{N-1}$. There remains
just one direction orthogonal to $\Sigma_{N-1}$ which we call $\psi$,
along which $W$ must vary. The superpotential equation then becomes
effectively one-dimensional leading to  a two-branched superpotential
as in the single-field case.  More generally, we can impose
that \eqref{b0} holds only on a sub-manifold $\Sigma_{N-p} \subset
\Sigma_{N-1}$, with $1\leqslant p\leqslant N$.

In order to write explicit solutions it will be convenient to choose
coordinates  such that $\Sigma_{N-p}$  corresponds to fixing $p$ of the
coordinates, which we label $\psi\equiv \psi^i$, $i=1,\dots,p$, to
their values $\psi_B^{i}$.  
The remaining coordinates will be denoted by $\L^\a$ with
$\a=1,\dots,N-p$. In other words, 
\eql{cb1}{
\phi=(\psi^1,\dots,\psi^p,\L^\a,\dots,\L^{N-p})\equiv(\psi,\L), 
}
\eql{cb2}{
	\Sigma_{N-p}
	=
	\le\{
		\phi \in\MM_\phi
		~|~
		\phi
		=
		(\psi^1_B,\dots,\psi^p_B,\L^\a,\dots,\L^{N-p})
		\equiv
		(\psi_B,\L)
		~,~
		{\p V\over \p\L ^\a}\Bigg|_{(\psi_B,\L)}=0\ri\}
~,
}
where the second condition is the equipotential condition. We will
also often use the notation $\phi_B=(\psi_B,\L)$.  In the vicinity of $\Sigma_{N-p}$, the superpotential can be expanded as
\eql{cb3}{
W(\phi)=W_B+\d W,\qquad  \d W\xrightarrow{\psi\to \psi_B} 0.
}
As we assume $V(\phi)$ is analytic,  we can write 
\begin{align}
V(\psi,\L)
=&V_B
+\sum_{i=1}^p{\p V\over \p\psi^i}\Bigg|_{(\psi_B,\L)}\d\psi^i
+\cO(\d\psi)^2
~
\label{cb4}
\end{align}
where $\d \psi=\psi-\psi_B$. Inserting the expansions (\ref{cb3}-\ref{cb4}) in the  superpotential
equation gives, to lowest order in $\delta \psi$ and $\delta W$,

\be \label{cb5}
\ha\p_r\d W\p^r\d W-{d\over 4(d-1)}\le(2W_B+\d W\ri)\d W = \sum_{i=1}^p{\p V\over \p\psi^i}\Bigg|_{(\psi_B,\L)}\d\psi^i
~,
\ee
where we have kept linear terms in $\delta \psi$ but up to {\em
  quadratic} terms in $\delta W$, because as we will see in a moment
some of these terms are of the same order as $\delta \psi$. Indeed,
suppose that $\delta W$ is of order $(||\delta \psi||)^\gamma$,
$\gamma > 0$. It follows that the three terms on the left hand side
of equation (\ref{cb5}) scale as 
\be\label{cb5-i}
W_B \delta W\sim  (||\delta
\psi||)^{\gamma}, \qquad  (\delta W)^2 \sim (\de_{\Lambda} \delta W)^2 \sim  (||\delta
\psi||)^{2\gamma}, \qquad (\de_{\psi} \delta W)^2 \sim  (||\delta
\psi||)^{2\gamma-2}. 
\ee
Of the three terms above, if  $\gamma <2$ the third  is the dominant
one for small $\delta\psi$, whereas the first term dominates if $\gamma >2$. In the latter case however it is
impossible to match the linear term on the right hand side. 
Therefore  it is the the third  term  in equation (\ref{cb5-i}) which dominates over the first two,
and we conclude that $\gamma = 3/2$ as in the single-field bounce case.
Equation (\ref{cb5}) can then be approximated  as, 
\be \label{Eik}
\p_r\d W\p^r\d W
 = 2\sum_{i=1}^p{\p V\over \p\psi^i}\Bigg|_{(\psi_B,\L)}\d\psi^i
~.
\ee
Notice that  equation \eqref{Eik} is  similar to the Eikonal equation of
geometrical optics, with  the left-hand side playing the role of the
square of the refractive index. Bounces are analogous to total
internal refraction (i.e.~mirages) off a region with vanishing refraction index.

There are many types of solutions to equation (\ref{Eik}). Those
which are closest to  the single-field case (\ref{BN6}) is
the following set of solutions,  differing by $2^p$ sign combinations:
\begin{align}
	\d W(\psi_B,\L)
	&
	=
	{2\over 3}
	\sum_{i=1}^p
	(-1)^{s_i}\sqrt{
			\le[{2\over \GG^{ii}}
			{\p V\over \p\psi^i}\ri]_{(\psi_B,\L)}
			\d\psi^i
	}
	\d \psi^i
	+\cO(\d\psi)^2~,
	~~
	s_i=0,1
	~.
	\label{cb6}
\end{align} 
We have written the solution so that the expression under
the square root is positive for both signs of $\de V/\de\psi_i$. 

Although all solutions  (\ref{cb6}) are valid close to the bounce surface
$\Sigma_p$,  only certain combinations can be glued together, as in
the single field case, to obtain regular flows  $\phi^r(u)$.  
To find which ways of gluing are consistent, we write the flow equation \eqref{floW} using equation  \eqref{cb6}
\begin{subequations}	\label{cb7}
\begin{align}
	&
	\dot \psi^i
	=
	(-1)^{s_i}
	\sqrt{
			\le[2{\GG^{ii}}
			{\p V\over \p\psi^i}\ri]_{(\psi_B,\L)}
			\d \psi^i
	}
	+\cO(\d\psi)~,
	~
	s_i=0,1
	~,
	\label{cb7a}
	\\
	&
	\dot \L^\a
	=
	\cO(\d\psi)^{3/2}
	~,~~
	\label{cb7b}
\end{align}
\end{subequations}
Integration of \eqref{cb7}, with the initial condition
$\phi^r(u_b)=(\psi_B,\L_B)$, leads to
\begin{subequations}	\label{cb8}
\begin{align}
	&
	\psi^i(u)
	=\psi^i_B
		+\ha
		\le[
			{\GG^{ii}\over 2}
			{\p V\over \p\psi^i}
		\ri]_{(\psi_B,\L)}
		(u-u_B)^2
	+\cO(u-u_B)^3~,
	\label{cb8a}
	\\
	&
	\L^\a(u)
	=\L^\a_B
	+\cO(u-u_B)^3 , 
	\label{cb8b}
	\end{align}
\end{subequations}
where in each direction we need to impose the consistency condition
\be \label{cb8c}
(-1)^{s_i}\le[
			{\p V\over \p\psi^i}
		\ri]_{(\psi_B,\L)}
	(u-u_B)>0 \, .
\ee
As in the single field case, the two sign choices $s_i=0,1$ in each direction
give the solution for $u>u_B$ and $u< u_B$, respectively. However, all fields
$\psi_i(u)$ must be in the same range ($u>u_B$ and $u< u_B$) for the
solution to be smooth. Therefore the condition (\ref{cb8c}) fixes all
but an overall sign choice, which out of the set (\ref{cb6})  leaves only two
solutions which glue consistently across the bounce, 
\begin{align} \label{cb9}
W(\psi,\L)
	=W_B
	\pm{2\over 3}
	\sum_{i=1}^p
	\text{sign}\le(\le[
			{\p V\over \p\psi^i}
		\ri]_{\psi_B}\ri)
	\sqrt{
			\le[{2\over \GG^{ii}}
			{\p V\over \p\psi^i}\ri]_{\psi_B}
			\d\psi^i
	}
	\d \psi^i
	+\cO(\d\psi)^2~,
\end{align}
The $\pm$ sign actually characterise two branches of the same multi-valued superpotential, as follows from the fact that both branches describe the same set of flows but for different ranges of $u$.

There exist other consistent solutions to \eqref{Eik} and many of them
share an interesting feature: they extend $\Sigma_{N-p}$ to a
higher-dimensional sub-manifold of $\Sigma_{N-1}$. For example, take  the following solution,
\begin{align}\label{cb10}
W(\psi,\L)
	=
	&
	W_B
	\pm {2\sqrt{2}\over 3}\Bigg\{
	\le[
		\sum_{j,k=p-n+1}^{p}\GG^{ij}{\p V\over \p\psi^j}{\p V\over \p\psi^k}
	\ri]_{(\psi_B,\L)}^{-1/2}
	\le(
		\sum_{l=p-n+1}^{p}
		\le[
			{\p V\over \p\psi^l}
		\ri]_{(\psi_B,\L)}
		\d \psi^l
	\ri)^{3/2}
	+
	\nonumber
	\\
	&
	+
	\sum_{i=1}^{p-n}
	\sqrt{
			\le[{1\over \GG^{ik}}
			{\p V\over \p\psi^i}\ri]_{(\psi_B,\L)}
			\d\psi^i
	}
	\d \psi^i
	\Bigg\}
	+\cO(\d\psi)^2~.
\end{align}
For this solution there is a bounce
at $\psi=\psi_B$  as before, but now this point belongs to a line of
bounces defined  by
\eql{cb11}{
\sum_{l=1}^{n\leqslant p}
		\le[
			{\p V\over \p\psi^l}
		\ri]_{(\psi_B,\L)}
		\d \psi^l=0~.
}
The bounce manifold is now $N-p+1$ dimensional. By a change of coordinates, we can recast the solution \eqref{cb10} in
the form \eqref{cb6} by defining one more  $\L$-coordinates for the
directions satisfying \eqref{cb11} and removing one
$\psi$-coordinate. 

Below we illustrate these features by considering complete bounces in
a simple two-scalar example.

\paragraph{Two-field example.} 
We set  $\phi=(\phi^1,\phi^2)$, and to make things simple we consider
a flat field space metric\footnote{By the use of Riemann normal
  coordinates, \eqref{RNor}, this assumption turns out to be a general
  at the order in $\phi^r$ in which we will work.}
$
\GG_{rs}=\d_{rs}.
$
The equipotentials are one-dimensional, therefore by our
general discussion we can have bounces at  isolated points or along
lines. We first assume that $\phi_B$ is an isolated  {\it point}, thus
all coordinates will be of the $\psi^i$ type.  We will see below that
the solutions in which bounces occurs on a line naturally appear  in this framework.

\begin{figure}[t]
\centering
\includegraphics[width=0.75\textwidth]{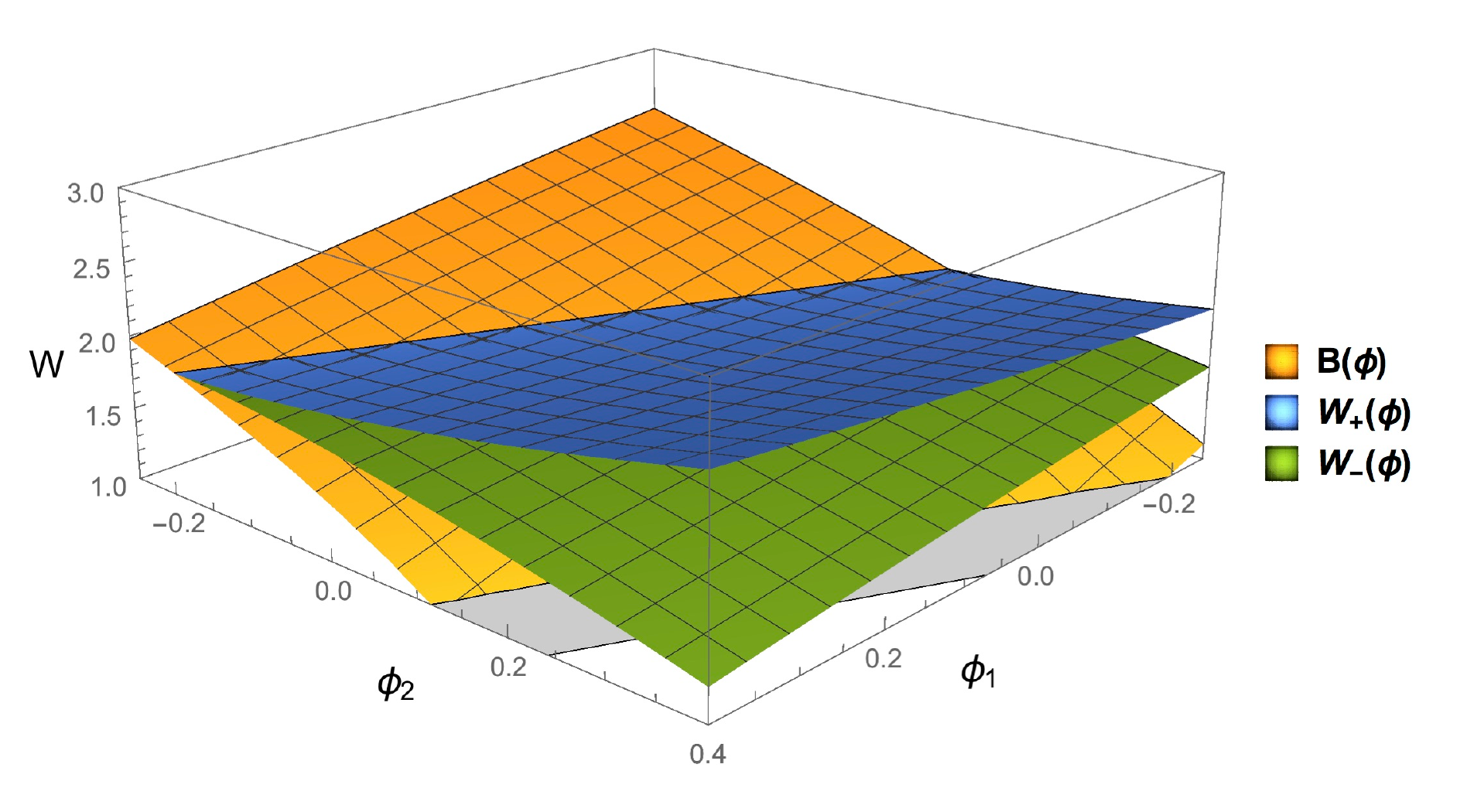}
\caption{Two fields - a complete bounce given, to leading order in $\phi-\phi_B$, by equation \protect\eqref{BDani1W}. The associated vector field is depicted in figures \protect\ref{fig:BDaniF}, with the lower (upper) branch in green (blue).}
\label{fig:BDani}
\end{figure}
	
For a generic complete bounce,
\eql{b14}{
 \partial_1 V(\phi_B)\neq 0, \qquad \partial_2 V(\phi_B)\neq 0.
}
Another way of writing the Eikonal equation \eqref{Eik} is:
\be
\vec{\nabla} (\delta W)  = \sqrt{2\sum_{r=1}^2\le.{\p_r V}\ri|_{B}\d\phi^r
} \vec{m}+\cO(\d\phi)^2
\equiv \sqrt{X}\vec{m}+\cO(\d\phi)^2
\label{b21}
\ee
for some, generally $\phi^1$ and $\phi^2$-dependent, unit vector $\vec{m}$. One can write this as $\vec{m} = d\vec{r}/{du}$ where $\vec{r}(u)$ is the `light ray' with $u$ an affine parameter along the path. An equivalent form of \eqref{b21} is
\be
\partial_1 \delta W(\phi) =  \sqrt{X}\cos(g(\phi)) \, \qquad \partial_2 \delta W(\phi) =  \sqrt{X}\sin(g(\phi))
\ee
where $g$ is a function satisfying, to leading order in $\delta \phi = (\phi-\phi_B)$, a constraint equation derived from $\p_{[1}\p_{2]}W=0$, namely
\be \label{BFR1}
(X\p_2 g+\le.\p_1V\ri|_B)\sin g+(X\p_1g-\le.\p_2V\ri|_B)\cos g
=0.
\ee

\begin{figure}[t]
\centering
\subfigure[Upper branch of $\p_rW$ from \protect\eqref{BDani1W}.]{
\includegraphics[width=0.45\textwidth]{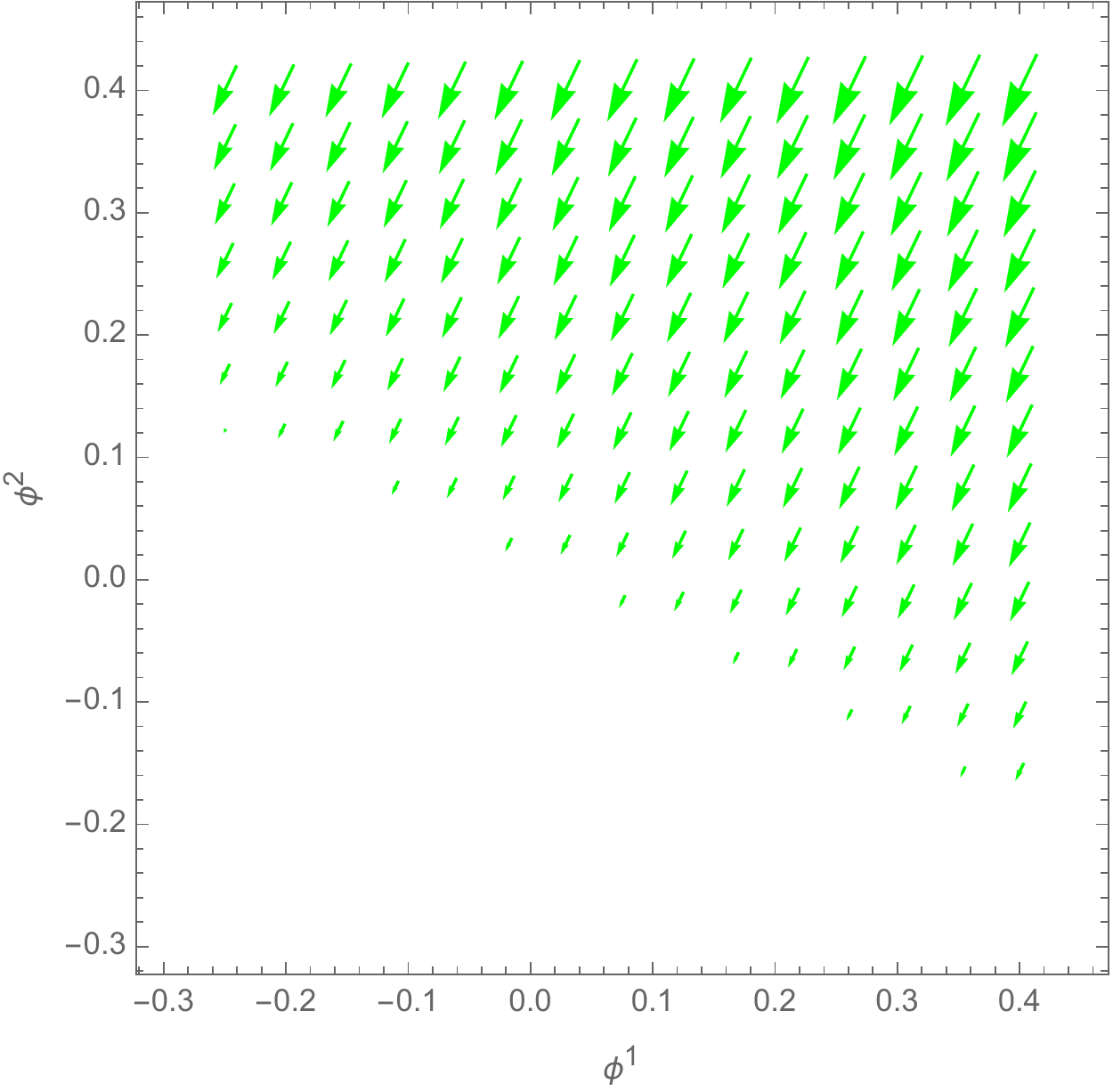}
}
\subfigure[Lower branch of $\p_rW$ from \protect\eqref{BDani1W}.]{
\includegraphics[width=0.45\textwidth]{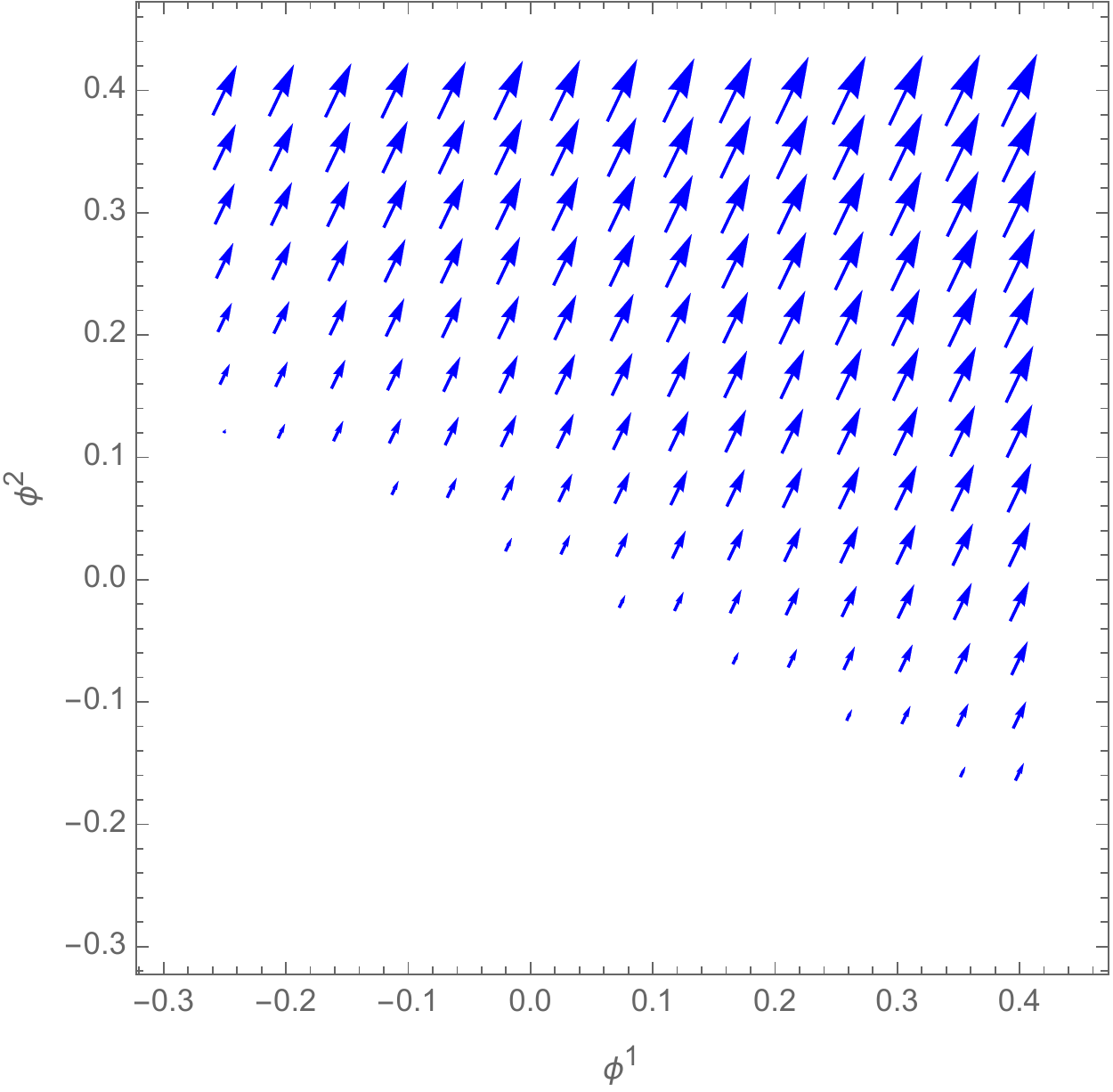}
}
\caption{The superpotential \protect\eqref{BDani1W} gives rise, to leading order in $(\phi^r-\phi^r_B)$, to the vector fields depicted above. In figure (a) the green arrows correspond to the gradient of the lower branch of $W(\phi)$ depicted in figure \protect\ref{fig:BDani}, the choice of a negative sign in \protect\eqref{BDani1W}. Figure (b) represents the gradient of the upper branch from figure \protect\ref{fig:BDani}, a positive sign in \protect\eqref{BDani1W}. Each flow line goes back along itself close to this complete bounce.}
\label{fig:BDaniF}
\end{figure}

One solution is
\eql{BDani1W}{
	W(\phi)
	=
	W_B
	\pm \frac{2\sqrt{2}}{3}\frac{\le[\partial_1 V( \phi_B )\d
          \phi^1+\partial_2 V( \phi_B )\d\phi^2\ri]^{3/2}}{\norm{ \p_s
            V|\le(\phi_B\ri)}}+\cO(\d\phi^r)^2 
}
corresponding to $\tan g = \le[\p_2V/\p_1V\ri]_B$. The solution \eqref{BDani1W} is represented in figure \ref{fig:BDani} and the gradients of its two branches are shown figures \ref{fig:BDaniF}. Notice that the bounce occurs along a line given by the solution to $X=0$ with $X$ defined in equation \eqref{b21}. It corresponds, therefore, to the maximal dimension of an equipotential in two dimensions.

\begin{figure}[t]
\centering
\includegraphics[width=0.75\textwidth]{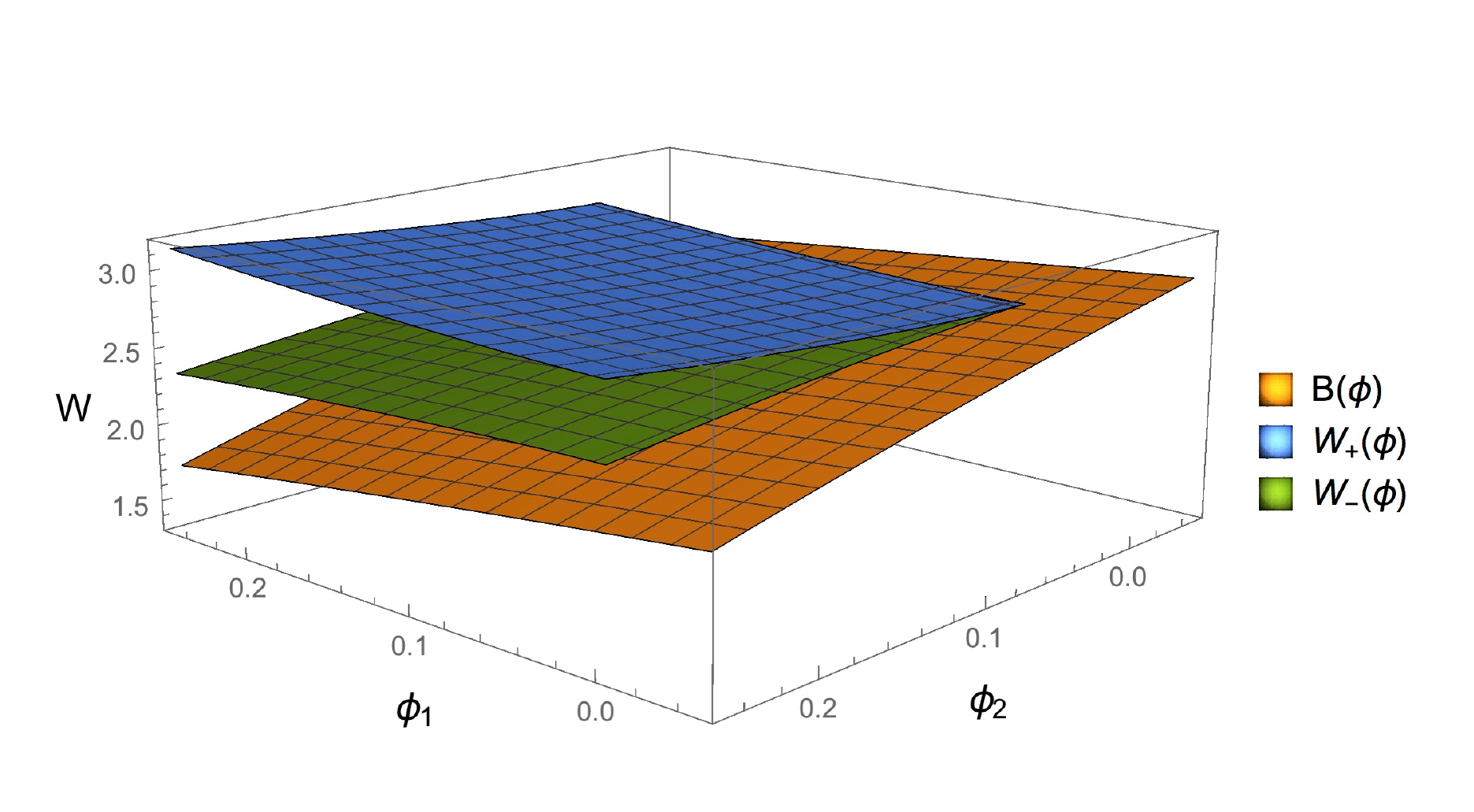}
\caption{Two fields - a complete bounce given, to leading order in $\d\phi^r=\phi^r-\phi^r_B$, by equation \protect\eqref{BDani2}. The two branches meet at a point.}
\label{f:BPoint}
\end{figure}

\begin{figure}[t]
\centering
\includegraphics[width=0.48\textwidth]{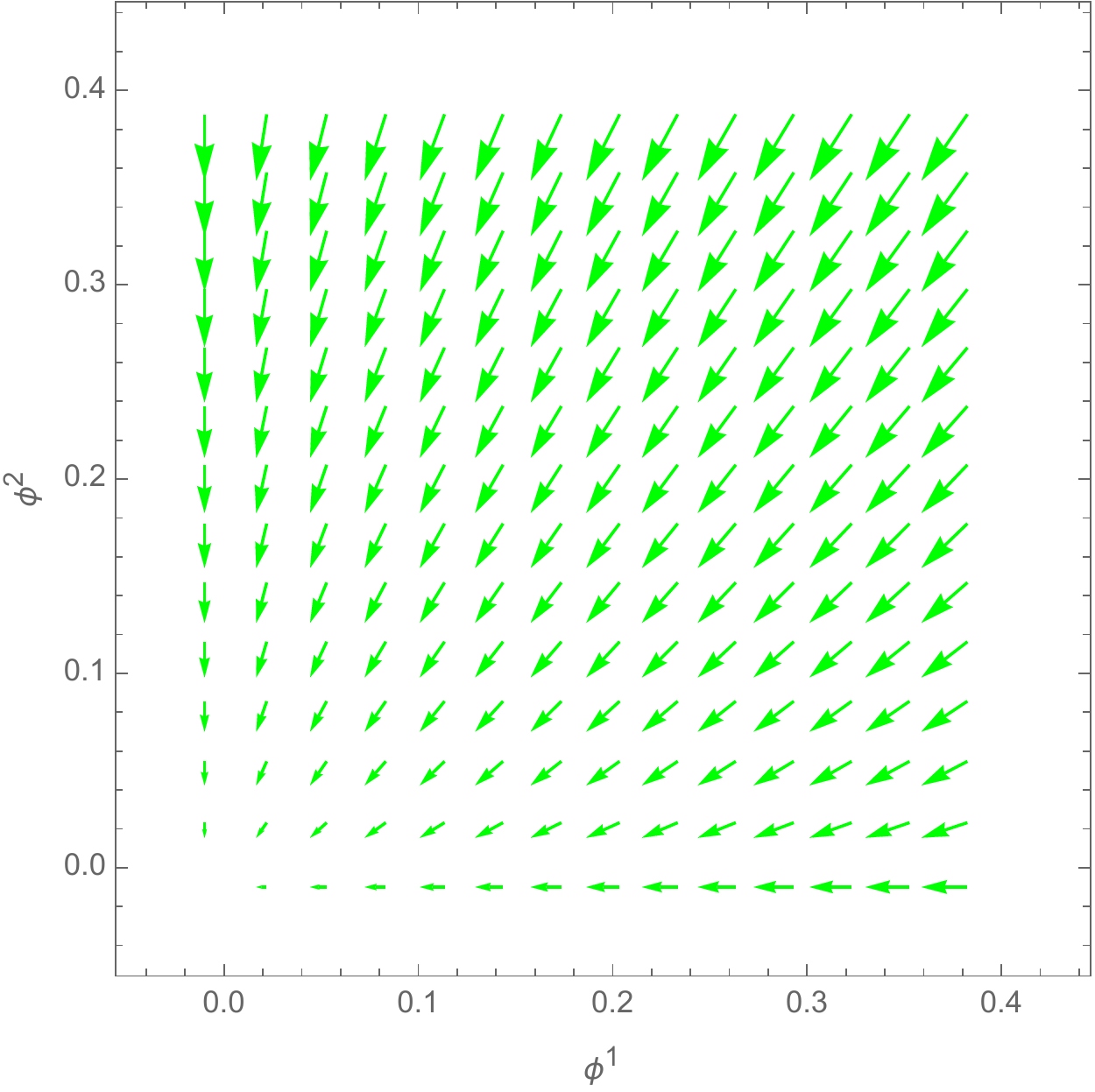}
~~
\includegraphics[width=0.48\textwidth]{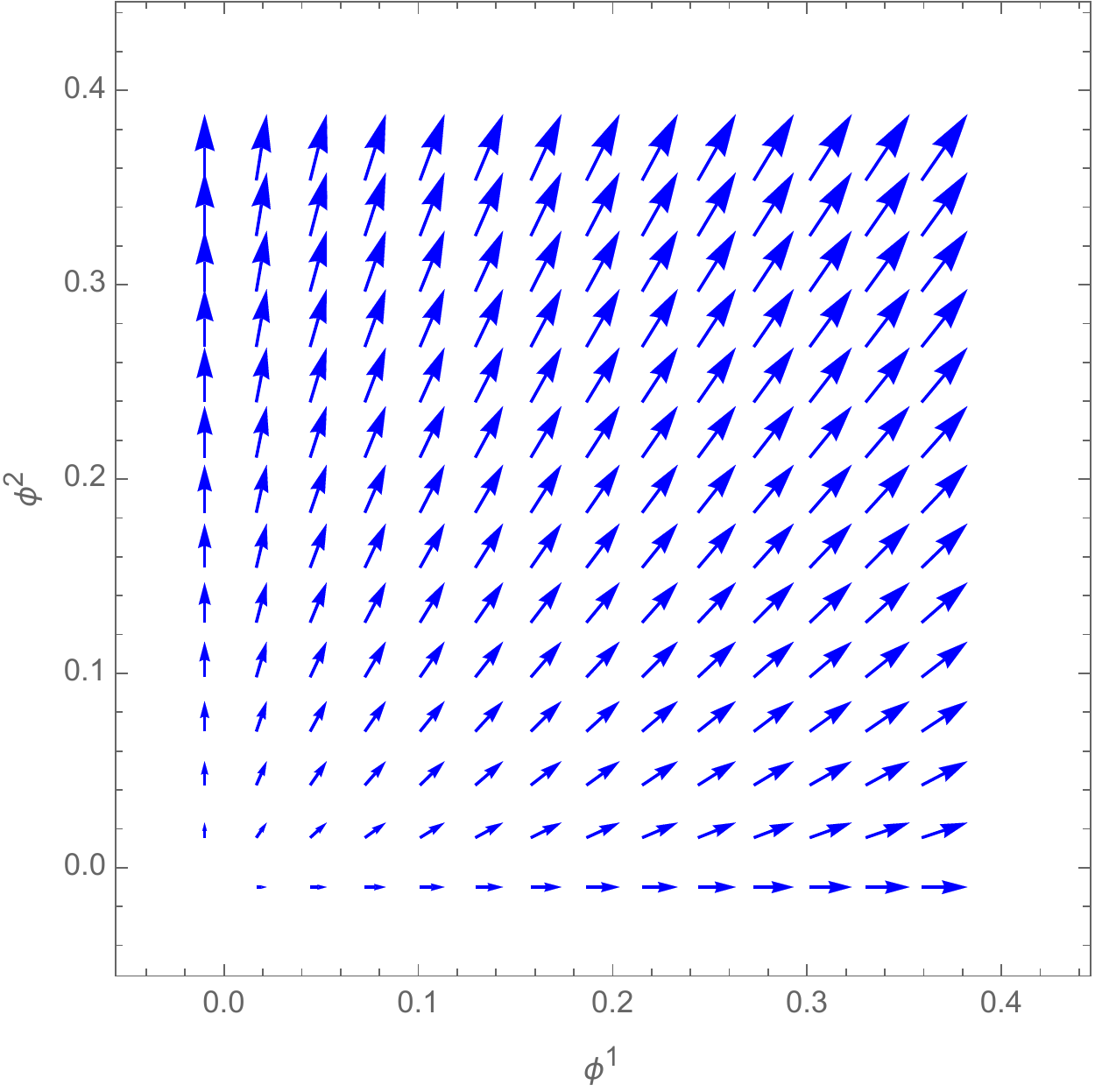}
\caption{The gradient of the superpotential given by equation \protect\eqref{BDani2} and depicted in figure \protect\ref{f:BPoint}. The velocity field focus on the point $\phi=\phi_B=0$. The fact that the $W(\phi)$ cannot be continued for negative $\phi^r-\phi^r_B$, as follows from \protect\eqref{BDani2} and from the signs of $\p_rV|_{\phi_B}$ for our potential, is not a problem since the paths this superpotential generates do not cross the lines $\d \phi^1=0$ and $\d \phi^2=0$.}
\label{f:BPoint2}
\end{figure}

Another solution to equation (\ref{BFR1}) is given by:
\eql{BDani2}{
W =W_B\pm{2\over 3}\sum_{r=1}^2
	\text{sign}\le(\le.\p_rV\ri|_{\phi_B}\ri) \d\phi^r\sqrt{2\le.\p_rV\ri|_{\phi_B}\d\phi^r}
+\cO(\d\phi^s)^2~,
}
where the choice of signs was fixed using the consistent solutions \eqref{cb9} and the $\pm$ sign is related to the range of $u$ in the expansion \eqref{cb8a} by
\eql{b17}{
\pm=\text{sign}\le[(u-u_B){\le.\p_rV\ri|_{\phi_B}}\ri].
}

While for the first solution, \eqref{BDani1W}, the bounce occurs on a
line, for this second solution, \eqref{BDani2}, the bounce takes place
{\it only} at the point $\phi= \phi_B $ because this is the only place
where $\norm{\p_r W}$ in equation (\ref{BDani2})  vanishes.

One apparent drawback of the expansion \eqref{BDani2} is that the the reality condition for the superpotential, $V_r\le(\phi^r-\phi^r_B\ri)>0$, defines a quarter of the plane with origin at the point $\phi_B$ and we cannot obtain the full solution which extends outside of this region from this expansion. However, from figure \ref{f:BPoint2} we observe that the velocity field is contained in this range, meaning that no solution of the form \eqref{BDani2} has flows escaping this region of the plane and an extension is, in practice, unnecessary.

Thinking in terms of the Eikonal equation \eqref{Eik}, complete
bounces can be seen as a total internal reflection on a meta-material
with a refractive index that vanishes linearly as we approach the
critical curve.

\subsubsection{Partial bounces}
\label{sss:pba}

Partial bounces occur on loci where some, but not  all field components have vanishing speed, in contrast with the complete bounces treated in subsection \ref{sss:cba}. A more precise definition is the following:\\

{\em A partial bounce corresponds to a change of branches of the
  superpotential occurring on a sub-manifold $\Sigma_{N-p}$ of
  $\MM_\phi$, such that $\norm{\p_rV|_{\Sigma_{N-p}}}$ is non-zero and
  the superpotential $W(\phi)$ does not equal $B(\phi)$ on
  ${\Sigma_{N-p}}$.}\label{defPb} \\

 At a partial bounce, as at a complete bounce, a single-valued superpotential fails $W(\phi)$ to describe locally geodesically complete solutions. A bounce is also characterised by the existence of a second superpotential which coincides with $W(\phi)$ on $\Sigma$ and which makes the flows locally geodesically complete. The superpotential should be seen as locally double-branched and its integral lines fail to define local coordinates around the bounce. 

For concreteness, we start by determining the behaviour of $W(\phi)$
around a partial bounce  occurring on a co-dimension one
sub-manifold. Bounces will no longer correspond to equipotentials and
we will see below which changes this implies with respect to the
complete bounces of subsection \ref{sss:cba}.  We compute the explicit
form of the holographic $\b$-function perturbatively near the
bounce. We show that the $\b$-function has two branches, as in the single-field case \cite{multibranch} and we highlight the differences from that case.

Our choice of coordinates on an open neighbourhood of a co-dimension one bouncing manifold, $\Sigma_{N-1}$, is such that
\eql{B13}{
\norm{\p W}^2
=
		\GG^{\psi\psi}\le({\p W\over \p\psi}\ri)^2
		+
		\GG^{ij}{\p W\over \p\L^i}{\p W\over \p\L^j}
		~,
}
where $\Sigma_{N-1}$ in these coordinates is characterised by $\psi=\psi_B$.
We can rewrite the superpotential equation \eqref{SuperP}, in these coordinates, as:
\eql{B14}{
{\p W\over \p\psi}
=\pm
\sqrt{{1\over \GG^{\psi\psi}}\le(2V(\psi,\L)
+\frac{d}{2(d-1)}W^2(\psi,\L)-\GG^{\a\b}{\p W\over \p\L^\a}{\p W\over \p\L^\b}\ri)
}
~,
}
with $\a=1,\dots,N-1$.
 Equation \eqref{B14} becomes degenerate for a given $\psi=\psi_B$ in the following if
\eql{B15}{
{\p W\over \p\psi}\Bigg|_{(\psi_B,\L)}=0
\sp
\GG_{\psi\psi}(\psi_B,\L)\neq0
\quad
\nd
\quad
\p_{\psi} V(\psi_B,\L)\neq0.
}
The condition that $\GG_{\psi\psi}$ is non-zero is, for metrics of the
form \eqref{B13} equivalent to the non-degeneracy of $\GG$. This
ensures that the vanishing of $\p_\psi W$ at $\Sigma_{N-1}$ is not a
consequence of ill-defined coordinates. By analogy with the one-dimensional case, we expect a pair of solutions to the superpotential equation that meet at $\Sigma_{N-1}$, corresponding to the $\pm$ sign in \eqref{B14}.

From \eqref{B14}, \eqref{B15} and our definition of a bounce, it follows that when $W(\phi)$ is restricted to $\Sigma_{N-1}$ it must satisfy:
\eql{B16}{
V_B(\L)\equiv V(\psi_B,\L)=
\ha\le[\GG^{ij}{\p W\over \p\L^i}{\p W\over \p\L^j}\ri]_{(\psi_B,\L)}
-\frac{d}{4(d-1)}W^2(\psi_B,\L)
~.
}
This is nothing but 
the $N-1$ dimensional superpotential equation \eqref{SuperP}.
Therefore, partial bounces occur along solution of the
lower-dimensional superpotential  equation restricted to the directions $\Lambda^i$.

Around $\psi=\psi_B$ we  can  expand $V$ and $W$ as follows:
\begin{subequations}
\label{B17}
\begin{align}
&V(\psi,\L)
=V_B(\L)+\d\psi{\p V\over \p\psi}\Bigg|_{(\psi_B,\L)} +\cO\le(\d\psi\ri)^2,
\quad
\text{with}
\quad
\d\psi\equiv\le(\psi-\psi_B\ri),
\label{B17a}
\\
&W(\psi,\L)
=W_B(\L)+\d W(\psi,\L)
\equiv
W_B(\L)+\d W.
\label{B17b}
\end{align}
\end{subequations}
where
\eql{B18}{
W_B(\L)\equiv W(\psi_B,\L)
}
is a solution to \eqref{B16}. 
%
Proceeding in the same way as for the complete bounce, we arrive
at the solution close to a partial bounce,
\begin{align}
	W(\psi,\L)
	=
	W(\psi_B,\L)
	\pm
	{2\over 3}\d \psi\sqrt{
	2\le[\GG_{\psi\psi}
			{\p V\over \p \psi}\ri]_{(\psi_B,\L)}
		\d \psi
		}
		+\cO(\d\psi)^2~.
		\label{B23}
\end{align}
The similarity with the single-field case and with the complete
bounces of the previous subsection  is manifest. The dependence of $W$
on $\d \psi$ has the same power  $3/2$ and the solutions are
restricted to $\psi<\psi_B$ {\it or} $\psi>\psi_B$, with two possible
signs defining branches. Both branches should again be seen as
belonging to the same superpotential in order to make the flow
$\phi^r(u)$ locally geodesically complete. The difference lies in the
fact that $W(\psi_B,\L)$ and the term under the square root are
functions of $\L^\a$ and not  constants. Therefore, along a partial
bounce the flow has non-vanishing speed along the surface $\Sigma_{N-1}$.  When all velocities vanish at the bounce we are back to the case of a complete bounce which we analysed in subsection \ref{sss:cba}.


As in the case of complete bounces, a partial bounce can also occur on a sub-manifold of $\MM_\phi$ with co-dimension $p>1$, $\Sigma_{N-p}$. The solutions \eqref{cb9} generalise to partial bounces by the simple replacement:
\eql{B28}{
W_B\to W_B(\L^\a)\quad\text{with}\quad i=1,\dots,p \quad\nd\quad\a=1,\dots,N-p~.
}
as long as $W_B(\L^\a)\equiv(\psi_B^i,\L^\a)$ solves \eqref{B16} and, of course, as long as the potential and the metric to be such that $\GG^{ir}\p_rV$ is non-zero on $\Sigma_{N-p}$. 

\begin{figure}[t]
\centering
\subfigure[]{
\includegraphics[width=0.6\textwidth]{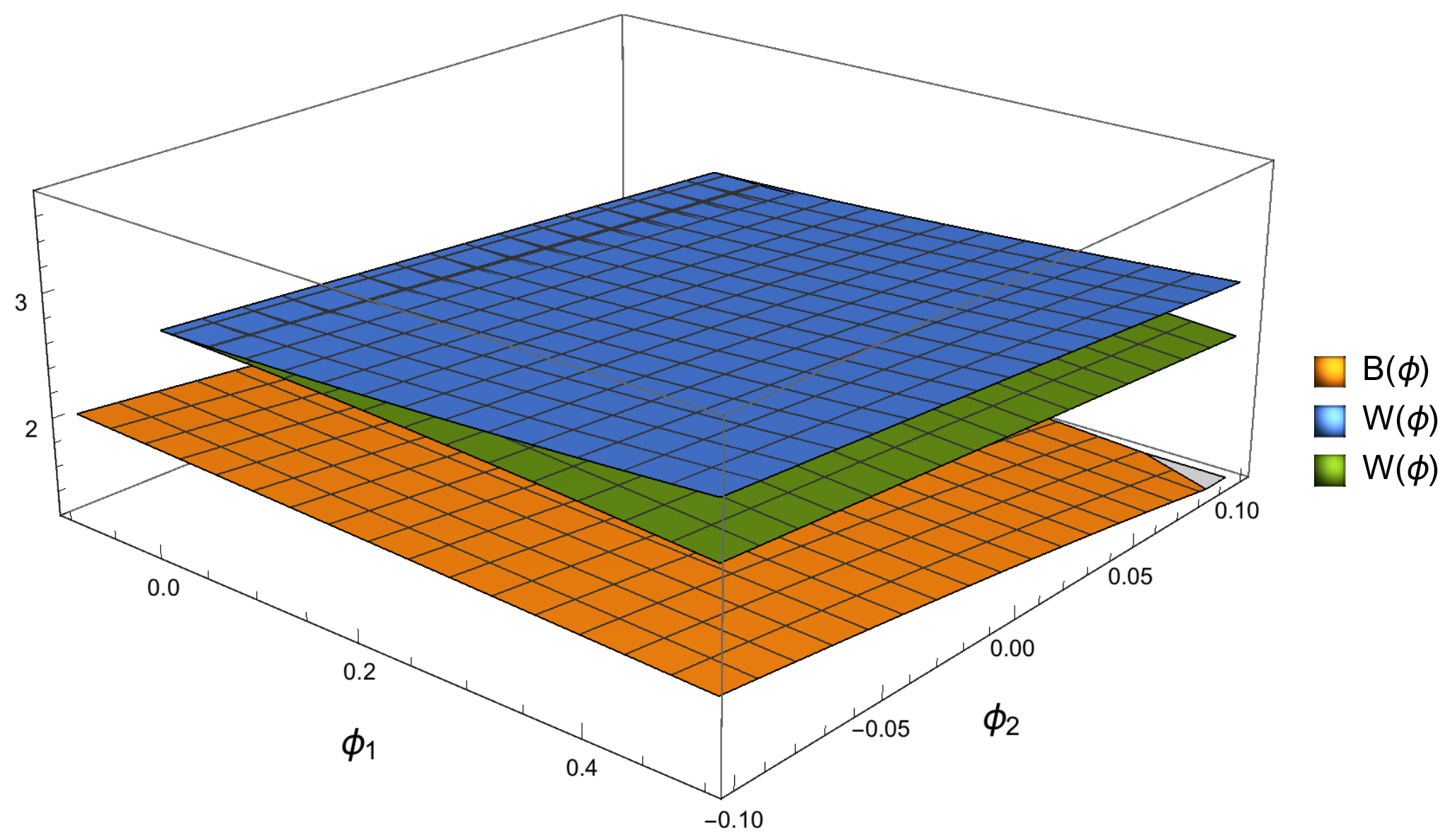}
}
\subfigure[]{
\includegraphics[width=0.35\textwidth]{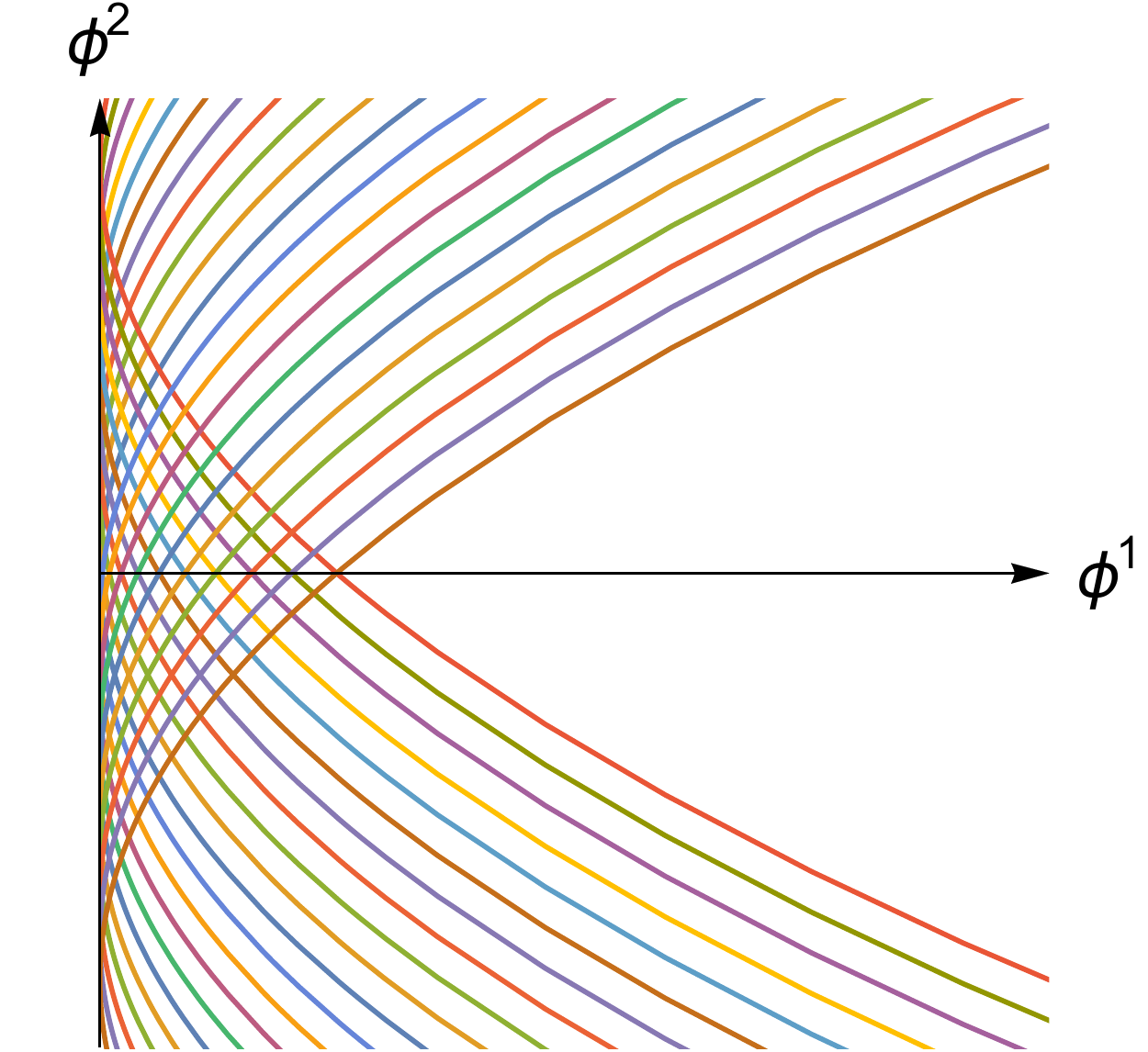}
}
\caption{Figure (a) represents in yellow the curve $B(\phi)$ and a two-branched superpotential $W$ in green and blue. $B(\phi)$ is associated with the potential $V(\phi)$ through equation \protect\eqref{B}. The plot is in the linear regime  \protect\eqref{B29} with parameters $V_0=-1$, $V_1=0.44$, $V_2=3.31$. The two branches of $W(\phi)$ are given by equations \protect\eqref{B30} and \protect\eqref{B31}. The upper branch of the superpotential corresponding to the $+$ sign in \eqref{B30} is in blue and the lower branch of $W$ is in green.
Figure (b) shows the flow lines around a bounce, from equation \protect\eqref{B32}, differing by the choice of $\phi^1_0$. Even though no individual line is self-intersecting, the ensemble of the flow lines gets superposed after the bounce.
} 
\label{f:pb}
\end{figure}

One important distinction needs to be made between partial and
complete bounces: a flow with a partial bounce can also be derived
from a superpotential with no bounces, as a consequence of the large
redundancy of the superpotential formalism in the multi-field case. On
the other hand, in a complete bounce the flow has to reach the
critical curve, and this   will be the case for any choice of
superpotential.

%

To conclude this section we give an example of a partial bounce in the
case of two scalar fields. 

\paragraph{Two-field Example.} Consider two scalars with a flat metric and a potential in the linear approximation such that:
\eql{B29}{
V(\phi^1,\phi^2)=C+V_1\phi^1+V_2\phi^2+\cO(\phi)^2, \quad V_1>0, ~V_2>0~.
}
For simplicity we will consider a pair of superpotentials of the form:
\eql{B30}{
W=W_0+W_2\phi^2\pm \sqrt{2V_1 \phi^1}\phi^1+\cO(\phi)^{5/2}
}
The expansion \eqref{B30} means that we have set $\le[\p^2W/ \p(\phi^2)^2\ri]_{(0,0)}$ to zero. Equation \eqref{B16} in this case is a single-field superpotential equation which has a continuum of solutions parametrised by an integration constant. Imposing $W(\phi)$ is of the form of \eqref{B30} already fixes this constant, fixing also the relation between $W_0$, $W_2$ and the coefficients appearing in \eqref{B29}:
\begin{subequations} \label{B31}
\begin{align}
&
W_0
=
\sqrt{\frac{2(d-1)}{d} \left(\sqrt{V_0^2+{2 (d-1)\over d} V_2^2}-V_0\right)}
\label{B31a}
\\
&
W_2
=
- \sqrt{{V_0\over V_2}+\sqrt{\le(V_0\over V_2\ri)^2+\frac{2 (d-1)}{d}}}\label{B31b}
\end{align}
\end{subequations}
The flows coming from the superpotential \eqref{B30} after solving \eqref{floW} are of the form:
\begin{subequations}\label{B32}
\begin{align}
&\phi^1(u)=\phi_0^1+{V_1\over 2}(u-u_B)^2+\cO(u-u_B)^3
\label{B32a}
\\
&\phi^2(u)=W_2(u-u_B)+\cO(u-u_B)^3
\label{B32b}
\end{align}
\end{subequations}

Figure \ref{f:pb}(a) shows the two branches of the superpotential \eqref{B30} plotted together with the potential \eqref{B29} for a specific choice of parameters. Figure \ref{f:pb}(b) displays the flows \eqref{B32} for different values of $\phi_0^1$. The flow lines return to themselves after the bounce, so we cannot use the flow lines to define one of the coordinates unless we choose a single branch of the superpotential. This figure suggests pursuing the analogy with geometrical optics which is suggested by the Eikonal equation \eqref{Eik} and the flow equation \eqref{floW}. As noted previously, a complete bounce is the analogous of a total reflection. What figure \ref{f:pb} suggests is that a partial bounce can be though of the analogue of a mirage. 

\section{Comments on the canonical formalism  for multi-field gravity}
\label{s:can}

The superpotential formalism used in the previous section is closely
related to the Hamilton-Jacobi formalism. In this section we will
turn to the general formulation in terms of the Hamilton-Jacobi
principal function to gain a more precise understanding why the  {\em
  gradient} flows assumption is generic. We will also argue that a
description  in terms of  non-gradient flows may be useful in some
situations, like in the presence of a global symmetry (this situation
however does not apply in the context of  holography). 



We begin by recalling a few facts from Hamilton-Jacobi (HJ) theory.
This provides  a procedure to map solutions $(q(t),p(t))$ of the equations of motion into constants $(\b,\a)$ which are functions of the initial conditions $(q_0,p_0)$ by means of a canonical transformation. The generating function of such a canonical transformation is called {\it Hamilton's principal function} and is usually denoted by $\SS(q^i,\a_i,t)$ where $\a_i$, $i=1,\dots,n$, are independent integration constants which define the new constant momenta. The canonical transformation defined by the principal function is such that:
\begin{subequations}\label{can8}
\begin{align}
p_i=&{\p \SS\over \p q^i},\label{Can8a}
\\
\b^i=&{\p \SS\over \p \a_i},\label{Can8b}
\\
H\le(q^i,{\p \SS\over \p q^i},t\ri)&+{\p \SS \over \p t}=0,\label{Can8c}
\end{align}
\end{subequations}
where $H(q,p,t)$ is the Hamiltonian of the mechanical system and \eqref{Can8c} is called the {\it Hamilton-Jacobi equation}.

Only derivatives of $\SS$ appear in \eqref{Can8c} implying that $\SS$
is determined only up to an additive constant. $\SS$ is a function of
$n+1$ variables and equation \eqref{Can8c} places a single constraint
on this function. To integrate \eqref{Can8c} it is therefore necessary
to specify an {\it integration function} of $n$ variables. However,
for $\SS$ to define a canonical transformation to new constant
coordinates $(\alpha,\beta)$ it is necessary that it contains $n$ {\it
  independent, non-additive} integration {\it constants}, showing that this formalism is highly redundant. The
remaining $n$ integration constants which are necessary to specify a
solution are obtained by integrating \eqref{Can8a}, where  the momenta
$p_i$ are written in terms of the velocities $\dot q^i$.

One very useful aspect of Hamilton-Jacobi theory is that it straightforwardly incorporates  the relationship between conserved quantities and symmetries. For example, when the Hamiltonian $H$ does not depend explicitly on time, energy is conserved and we can write:
\eql{Se}{
\SS(q,P,t)=\WW(q,P)-Et~.
}
The function $\WW(q,P)$ is called Hamilton's characteristic function.
When \eqref{Se} holds, equation \eqref{Can8c} takes the following form:
\eql{Can11}{H\le(q^i,{\p \WW\over \p q^i}\ri)=E.}
In other words, principal functions of the form \eqref{Se} group
solutions to the equations of motion which have the same energy. 
Similarly, when one of the momenta $p_{\hat i}$ is conserved and has
the specific value $\widehat P_{\hat i}$, by integrating \eqref{Can8a}
one can isolate the dependence of $\SS$ on $q^{\hat i}$:
\be
\SS(q^i,p_i)  = \widehat{\SS} + {\widehat P}_{\hat i} q^{\hat i} 
\ee
where $\widehat{\SS}$ is independent of $q^{\hat i}$. 

\subsection{Gradient flows revisited}
\label{ss:grad}

We will  now implement the HJ formalism outlined above in the case of
gravity coupled to multiple scalars. This is a well known procedure  \cite{SkenderisTownsend,PapaCan} and here  we want to focus on the question of what set
apart gradient flows with respect to the more general non-gradient
flows mentioned in section \ref{ss:1of}. More specifically, we would like to answer the question of when
is it necessary to consider the non-gradient case. As we will see,
this question is related to the {\em separability} of the
$A$-dependence in the HJ principal function $\SS(A,\phi^r)$. 

An effective Lagrangian which leads to the equations of motion \eqref{gen3} is given by:
\eql{lag0}{
	L(A,\phi^r,\dot A,\dot \phi^r,N)
	=
	N e^{dA}
	\le[
		{d(d-1)\over N^2}\dot A^2
		-{1\over N^2} \GG_{rs}\dot\phi^r\dot\phi^s
		-V(\phi)
	\ri]~.
}
where $\dot ~={d\over d u}$: the ``time" for the evolution of the scalars and the scale factor is the holographic coordinate. In \eqref{lag0} the variable $N$ does not have a $u$-derivative and represents a constraint.
The equations of motion following from \eqref{lag0} are:
\begin{subequations} \label{lag1}
\begin{align}
	&
	{d(d-1)\over N^2}\dot A^2
		-{1\over 2N^2} \GG_{rs}\dot\phi^r\dot\phi^s
		+V(\phi)=0~,\label{lag1a}
	\\
	&
			\le[
				{{\GG_{rs}}\over N}
				\le(
					\ddot \phi^r
					+\WG^s_{pq}\dot\phi^p\dot\phi^q
					+d\dot A\dot \phi^s
					-{\dot N\over N}\dot \phi^s
				\ri)
				 -N \p_rV
			\ri]
		=0~,\label{lag1b}
	\\
	&
			\le[
				{d(d-1)\over N}\le(2\ddot A+d\dot A^2-2\dot A{\dot N\over N}\ri)
				+{d\over 2N}\dot \phi^2
				+d ~N ~V
			\ri]
			=0~.\label{lag1c}
\end{align}
\end{subequations}
Because $N$ is not dynamical, we can set $N=1$,  in which case
equations \eqref{lag1} become the original equations of motion
\eqref{gen3}.  The  canonical momenta are
\begin{subequations}\label{lag4}
\begin{align}
	&p_s=
	-e^{dA}\GG_{rs}\dot \phi^r
	~,
	\label{lag4a}
	\\
	&p_A=
	e^{dA}{2d(d-1)}\dot A
	~
	.
	\label{lag4b}
	\end{align}
\end{subequations}
and the Hamiltonian  is given by
\eql{lag5}{
	H(A,\phi^r,p_A,p_r)
	=
	e^{-dA}
	\le(
		{p_A^2\over 4d(d-1)}
		-\ha p_r p^r
	\ri)
	+e^{dA}V(\phi)~.
}
Written in terms of the $H$, equation (\ref{lag1a}) is just the
Hamiltonian constraint, 
\eql{lag2}{
H=\dot \phi^r p_r+\dot A p_A-L =  0.
}
This  reduces the number of independent integration constants from
$2N+2$ to $2N+1$ (the dynamics is constrained to stay on the
zero-energy surface). 

The HJ principal function $\SS$ is such that 
\eql{lag6}{
p_A={\p\SS\over \p A},
\quad
p_r={\p\SS\over \p \phi^r}.
}
and it satisfies the HJ equation following from the Hamiltonian
constraint (\ref{lag2}): 
\eql{lag7}{
	e^{-dA}
	\le[
		{1\over 4d(d-1)}\le({\p\SS\over \p A}\ri)^2
		-\ha \GG^{rs}{\p\SS\over \p \phi^r}{\p\SS\over \p \phi^s}
	\ri]
	+e^{dA}V(\phi)
	=0
}
Comparing \eqref{lag7} with \eqref{Se} and \eqref{Can8c} we see that
Hamilton's principal function $\SS$ is the same as  the characteristic
function $\WW$, since $E=0$. This also means that we need $N$
independent integration constants in $\SS$  
in order to have a well defined map from $(q,p)$ to the initial
conditions $\SS$. Together with the $N+1$ additional ones coming
from the first order flow equations (\ref{lag6}), this makes up the needed
$2N+1$ integration constants.  


A {\it special class} of Hamilton's principal functions $\SS$ is of the following separable form between $A$ and $\phi^r$,
\eql{hj1}{
\SS=-e^{dA}W(\phi)~.
}
With this ansatz,  equation  \eqref{lag7} reduces to  the
superpotential equation \eqref{SuperP} and  equations \eqref{lag4}
become the flow equations \eqref{floW} and \eqref{floA}. Moreover, the
flow is automatically gradient, as $\pi_r = \de_r W$. 

The converse statement is slightly more subtle:  given a solution which
can be written in terms of a gradient flow, it does not follow that
the associated principal function has the  separable form
(\ref{hj1}). On the other hand, one can show that in this case {\em there
exists } at least one  separable solution to Hamiton-Jacobi equation
(\ref{lag7}). Moreover, all non-separable solutions coincide with the
separable one on-shell. 
This point is developed in more detail in Appendix
\ref{a:Non}, where we also show how to recover non-gradient flows of
the form (\ref{v1}) with non-zero curl (\ref{v5}-\ref{v5-i})   from a
non-separable solution $\SS(A,\phi^r)$. 

We can conclude that,  if  solutions  of equation (\ref{lag7})  of the separable
form (\ref{hj1})  contain enough independent  integration constants,
one can be sure that any solution of the full system can be written in
the form of a gradient flow.  In the context of  holography, when 
solutions  connect to a fixed point in the UV which is  a local {\em
  maximum} of the potential, this is indeed
the case, as we have seen in Section \ref{s:hol}: close to the UV
fixed point,  the set of solutions 
with asymptotic form given in  equation (\ref{special}) provides all 
integration constants to reproduce  any near-boundary asymptotics of
the form (\ref{h7}). Moreover, if the solution goes through a bounce,
there is a unique way to connect the two branches into a regular
solutions, meaning that one does not introduce possible non-gradient
flows at bounces. 

It has also been noted (see
e.g. \cite{Sonner:2007cp,Dorronsoro:2016pin}) that, away from a bounce,  locally {\em any}
solution can be embedded in a non-gradient flow with an appropriate
superpotential (which however may not be globally defined on field
space).  We show how to perform this local reconstruction of the
superpotential to lowest order in  In Appendix \ref{app:local}.

It follows from the discussion above that, at least  in holography, one may safely forget non-gradient
flows. However, there are situations in which using a non-separable HJ
function may actually be convenient: this is the case when there are
global symmetries on the scalar manifold and we may decide to classify
solutions in terms of the corresponding conserved quantities. This
will be the subject of the next subsection.



\subsection{Symmetries vs. gradient flows} 
\label{ss:sym}

As we have discussed in the previous section, locally all flows can be
 put in a gradient form, and in holography gradient flows are enough
 to generate all bulk solutions which connect to a maximum of the
 potential in the UV. However, in this section we argue that in  some contexts it can be useful to
 consider {\it  non-separable} principal functions, giving rise to
 non-gradient flows, in particular  when symmetries
of the scalar sector are present. 

Consider a system in our setup characterised by an $O(2)$ isometry in field space with Killing vector field $k^s(\phi)$. We associate a coordinate $\T$ to the motion along the integral lines of this vector field. For an infinitesimal transformation that takes $\T$ to $\T+\d \T$ we have
\eql{curl0}{
	\d_\t\phi^s:= k^s(\phi)\d\T.
}
Invariance of the potential translates into:
\eql{curl20}{
	\d_\T V(\phi)=\d \T~ k^s\p_sV(\phi)=0.
}

As a result of the symmetry \eqref{curl20} and the fact that it is an
isometry, the momentum conjugate to $\T$, $p_\T$, is a constant of the
motion:
\eql{curl1}{p_\T=-e^{dA} \GG_{\T s}(\phi^\a)\dot \phi^s,
\qquad
\dot p_\T=0,
\quad
\a=1,\dots,N-1,\quad\phi^N=\T}
It is therefore, possible to write a Hamilton's principal function that groups flows with the same angular momentum $p_\T=L$ as follows:
\eql{curl2}{
\SS=\SS_L(A,\phi^1,\dots,\phi^{N-1})+L\T
.
}
The principal function \eqref{curl2} is clearly non-factorisable but
is advantageous in practice as it allows to group solutions according the
value of conserved charge $L$. 
%
%
From (\ref{curl2}) we define $\widehat{W}_L(\phi^\a,A)$ by
\eql{curl7}{
\widehat{W}_L(\phi^\a,A)\equiv -e^{-dA}\le(\SS_L(\phi^\a,A)+L\Theta\ri), \quad \a=1,\dots,N-1, \quad \phi^N=\Theta~.
}
 
The canonical momenta associated with $A$ and $\phi^r$ can be obtained from the principal function \eqref{curl2} by the use of equation \eqref{lag6}. 
However, in order to know if a flow is gradient {\it on the scalar manifold}, it is necessary to eliminate the $A$-dependence in the momenta. This can be achieved by finding a function $\AA(\phi)$ which equals $A(u)$ on flows and, as explained in appendix \ref{a:Non}, is a solution to the differential equation:
\eql{curl8}{
\GG^{rs}\le[\p_rS(\phi,A)\ri]_{A=\AA(\phi)}\p_s\AA={1\over 2d(d-1)}\p_A\SS(\phi,A)\big|_{A=\AA(\phi)}
~.
} 
Once a solution $\AA(\phi)$ is provided, we can use it to ``project"
the momenta onto the scalar manifold. It is convenient to define 
\begin{subequations}\label{curl11}
\begin{align}
&W(\phi):=-d^{-1}e^{-d\AA(\phi)}\le[\p_AS(\phi,A)\ri]_\AA~,
\label{curl11a}
\\
&\pi_r(\phi):=-e^{-d\AA(\phi)}\le[\p_rS(\phi,A)\ri]_\AA~.
\label{curl11c}
\end{align}
\end{subequations}
because when equations \eqref{curl11} are combined with \eqref{lag6} we obtain:
\begin{subequations}\label{curl12}
\begin{align}
	&\dot A
	=-\frac{W(\phi)}{2(d-1)}
	~,
	\label{curl12a}
	\\
	&\dot \phi^r
	=\GG^{rs}\pi_s=\pi^r
	~
	.
	\label{curl12b}
	\end{align}
\end{subequations}
which are precisely equations \eqref{v2} and \eqref{v1}. By projecting
in the same way the Klein-Gordon equations and the Einstein equations which are derived from \eqref{lag4}, \eqref{lag6} and \eqref{lag7} one obtains, as derived in appendix \ref{a:Non}, the following equations
\eql{recall}{
	\pi_p
	=
	\p_p W
	+
	{ 2(d-1)\over d~W} 
		\pi^s\le(\WN_s \pi_p-\WN_p\pi_s\ri).
}
which are nothing but equations \eqref{v5}. The term in parenthesis in \eqref{recall} is the curl of the vector field $\pi^r$, which we can compute from $\SS(A,\phi)$ after solving \eqref{curl8}. In other words, {\it we can map the principal function $\SS(A,\phi)$ to the velocity field $\pi_r(\phi)$ it generates on field space and determine the curl of $\pi^r$}.

For the special case in which one wants to {\it classify solutions in terms of conserved charges}, equation \eqref{recall} is necessary if working in field-space only.  An example is provided by the principal function \eqref{curl2} which is associated with the conserved charge $L$. The symmetries of the problem imply that a solution $\AA(\phi)$ to \eqref{curl8} can be taken to be independent of $\T$. It is then possible to define $W_L(\phi)$, the function $W(\phi)$ associated with a given value of the conserved charge $L$ from \eqref{curl7} and \eqref{curl11a}.
\eql{}{
W_L(\phi^\a):=\widetilde W_L(\phi^\a,\AA(\phi^\a)), \qquad \a=1,\dots, N-1~,\quad \phi^N=\T.
}
Once this is done, the component of the velocity field associated with the conserved charge, $\pi_\T$, satisfies
\eql{curl9}{
{\p^\a W_L}
{\p_\a \pi_\T}
={d\over 2(d-1)}W_L~ \pi_\T~.
}
If $W_L$ is non-zero, equation \eqref{curl9} implies that a non-vanishing $\pi_\T$ will necessarily have a $\phi^\a$ dependence. We also know from \eqref{curl2} and the $\T$-independence of $\AA(\phi^\a)$ that the other other components of the velocity field, $\pi_\a$, are also $\T$-independent. As a consequence, the velocity field will have a non-zero curl
\eql{curl10}{
\FF_{\a\T}\equiv \p_\a \pi_\T-\p_\T \pi_\a=\p_\a \pi_\T \neq0.
}
The principal function \eqref{curl2} leads to a \emph{non-gradient velocity field} over $\MM_\phi$ 

It is important to remark that the conservation of angular momentum does not forbid the existence of factorisable principal functions, i.e.~ of the form \eqref{hj1}. A factorisable solution corresponds in this case to a family of flows in which different flows have different angular momenta, while in the non-separable solutions of the form \eqref{curl2} all the flows have the same angular momentum $L$.

\paragraph{UV behaviour.} It is instructive to make the above
discussion explicit close  to a UV fixed point  holographic RG flows.

Consider the case of two scalar fields $(\phi_1,\phi_2)$ with flat field-metric and a potential
$V(R)$, which has a maximum
at $R=0$ (which will serve as a UV fixed point),  where 
\eql{sn0-s}{
\begin{cases}
\phi^1=R \cos \T
\\
\phi^2=R \sin \T
\end{cases}
}
The metric components are:
\eql{sn1-s}{\GG_{RR}=1,\quad\GG_{\T\T}=R^2,\quad\GG_{R\T}=0.}
We are interested in constructing paths which define Archimedean spirals:
\eql{sn2-s}{R=\a~(\T-\T_0),\qquad \T_0\in[0,2\pi)~.}
  
Close to $R=0$ the solution has the expansion (\ref{h7}) with
$\Delta_1 = \Delta_2 >0 $ and $u \to -\infty$. 
If we now write the conserved angular momentum using \eqref{curl0},  \eqref{sn0-s} and \eqref{sn1-s} we obtain:
\begin{align}
\label{as4}
L
=\phi^1 p_2-\phi^2 p_1
=e^{d A(u)}\le(\phi^1 \dot \phi^2-\phi^2 \dot \phi^1\ri)
\end{align}
Using  the asymptotic expansions (\ref{h7}) of $\phi^r$ and $A$ we obtain
\eql{as5}{
L
=
{(2\D^+-d)}\le(\phi^1_-  \phi^2_+-\phi^2_- \phi^1_+\ri),
}
Therefore, fixing $L$ places a constraint relating the sources
$\phi^r_-$ and the VEVs $\phi^r_+$ in the UV, in contrast with the usual
procedure of keeping {\it only} the sources fixed. 

Close to the UV we can also obtain an approximate expression for the
non-separable Hamilton principal function with fixed $L$. The
computation is presented in Appendix \ref{a:nonsep}. The final
result takes the form of an expansion in $e^{-dA}$, which tends to
zero in the UV,  
\be
\SS_L(A,R,\Theta) = -e^{dA} W_0(R) + L \Theta + e^{-dA}W_2(L,R) +
\cO\le(e^{-2dA}\ri), 
\ee
where $W(R)$ satisfies a radial superpotential equation which is
{\em independent} of $L$. The last term is proportional to $L^2$ and its
explicit form is given in equation (\ref{L5}). 

Notice that $L$ enters at sub-leading order in the  expansion, and in
the UV $\SS$ is again separable. This is consistent with the fact that
$L$ is proportional to the VEV terms in the solution. 

\subsection{Back to holography: Gauging global symmetries} \label{ss:gauge}

From a gravitational perspective, the solutions in the previous
sub-section coming from a non-gradient velocity field over the scalar
manifold are perfectly acceptable. However, as we will explain below,
in holography these solutions are unphysical, since  all  symmetries
of the scalar manifolds must be gauged. As we will see, this implies
that Poincar\'e-invariant solutions have necessarily zero charge. 

It has been argued  that quantum gravity does not allowed  for exact
global symmetries (see e.g. \cite{Banks:1988yz}). In the context of the
AdS/CFT correspondence, this argument can be made  precise, since
in gauge/gravity duality global symmetries on the boundary require gauge-fields in the bulk \cite{Aharony:1999ti}. 
In our setup, this translates into the requirement that any isometry of $\GG(\phi)$ which leaves the potential $V(\phi)$ invariant should be gauged.

We proceed now to gauge the $O(2)$ isometry of a two-dimensional and flat scalar manifold.

In cartesian coordinates the action of the $O(2)$ isometry on fields $\phi^r$ is linear and its infinitesimal form is given by:
\eql{gsn0}{
	\d_\T \phi^s:=\T \e^{sp}\phi_p.
}
where, $\e^{sp}$ is the Levi-Civita symbol. When the potential is a function only of $\abs{\phi}$, the transformation \eqref{gsn0} leaves the action \eqref{gen0} invariant. In order to gauge the transformation \eqref{gsn0}, we introduce the gauge field $A_a(x)$ which transforms as:
\eql{gsn1}{
\d_\T A_a=\p_a\T
}
and is minimally coupled to $\phi$. Our convention for the covariant derivative is:
\eql{gsn2}{
D_a\phi^s(x)=\p_a\phi^s(x)-A_a(x) \e^{sp}\phi_p.
}
The associated field strength is denoted by $F_{ab}$, and the general
form of the action is
\begin{subequations}
\label{gsn4}
\begin{align}
&S=M^{d-1} \int_{\MM} \dd^{d+1}x \sqrt{-g} \left[
		{1\over 2\kappa}R
-\frac{1}{2}\GG_{ps}D_a \phi^pD^a \phi^s -V(\phi^q)
-{Z(\phi)\over 4}F_{ab}F^{ab}
 \right]+S_{GH}
~,\label{gsn4a}
\end{align}
\end{subequations}
where the function $Z(\phi)$ gives an extra coupling of the scalars to
the gauge fields.

The equations of motion following from the action \eqref{gsn4} 
are:
\begin{subequations}\label{gsn4bis}
\begin{align}
	&Z\nabla_bF^{ba}
	+
	F^{ba}
	\p_b\phi^s
	\p_sZ
	-
	\e_{ps}\phi^p
	D^a\phi^s
	=
	0	 ~, \label{gsn4bisa}
\\
&	\DD_a (D^a\phi^s)
	-\GG^{sp}\le(
	\p_pV
	+{1\over 4}F_{ab}F^{ab}\p_pZ
	\ri)=0
	 ~, \label{gsn4bisb}
\\
&R_{ab}
	-\ha g_{ab}R
	=
	\kappa \Big[
		 \GG_{pq}
		 	\le(
				D_a\phi^pD_b\phi^q
				-g_{ab}
					\ha D_c\phi^pD^c\phi^q
					\ri)
			+
	\nonumber\\
	&
	\qquad\qquad\qquad\qquad
	g_{ab}V(\phi)
	+
	 Z(\phi)
	\gk_{IJ}
		\le(
			F^{I}_{ac} F^{J\cdot c}_{~b\cdot}
			-{g_{ab}\over 4}
			F^{I}_{cd}F^{Jcd}
		\ri)
	\Big].\label{gsn4bisc}
\end{align}
\end{subequations}
where $\DD_a$ is a gauge and field-space covariant derivative.
The most general ansatz preserving $d$-dimensional Poincar\'e
invariance is
\begin{subequations}\label{gsn5}
\begin{align}
&\dd s^2=\dd u^2+e^{2A(u)}\emn \dd x^\m\dd x^\n~,\label{gsn5a}
\\
&\phi^s=\phi^s(u)~,\label{gsn5b}
\\
&A_a(u)=\d_a^uA_u(u) \label{gsn5c}
\end{align}
\end{subequations}
The independent equations of motion \eqref{gsn4bis} for the ansatz \eqref{gsn5} are:
\begin{subequations}\label{gsn6}
\begin{align}
	0
	=
	&
	\phi^p
	\e_{ps}
	D_u\phi^s
	,
\label{gsn6a}
\\
	0
	=
	&
	\le[\d^p_q(\p_u+ d \dot A) -A_u \e^{p}_{~q}\ri]D_u\phi^q
	+
	\WG_{sq}^p D_u\phi^sD_u\phi^q
	-\GG^{pq}
	\p_qV
	,
\label{gsn6b}
\\
	0
	=
	&{d(d-1)\over2\kappa}\dot A^2
	- \ha
		(\dot\phi-A_u\phi)^2
			+V(\phi)
	.
\label{gsn6c}
	\end{align}
\end{subequations}
Equations \eqref{gsn6a} means that the angular component of the covariant derivative is zero for the pure gauge configuration \eqref{gsn5c} and it can be rewritten as:
\eql{}{
	\dot \T(u)=A_u(u).
}
This implies that if we gauge the $O(2)$ isometry, any motion along
the angular direction becomes a gauge artefact. Equivalently, we can
go to the unitary gauge $A=0$ and find that Gauss's law (\ref{gsn6a})
implies $L=0$.

\section{Conclusion} \label{s:end}
In this work we have performed a systematic analysis of Einstein
gravity coupled to $N$ scalar fields, using the first order formalism
of flow equations.  Although our results were mostly framed
in the language of the gauge/gravity duality, they can also be used in
the context of cosmology if one trades the holographic coordinate for
time and flips a few signs. 
 
Our analysis shows that in holography, when bulk solutions have a
maximum of the potential to connect to in the UV, one can always write
the solution as gradient flows coming from a
superpotential  function. The latter may have branch points
corresponding to bounces, when one or more of the coordinates on field
space invert their flow direction. We have written  the general
solution close to an extremum of the bulk  potential in terms a
universal analytic superpotential plus  and a
set of continuous sub-leading deformations, which carry the integration
constants which ultimately determine the fate of the solution in the
infrared. 

As in the single-field case, an appropriate regularity condition in
the IR at minima of the scalar potential 
determines the full superpotential  completely. 

We have extended to the multi-field case the analysis of bouncing
solutions, for which the superpotential becomes multi-branched. We
have shown that complete bounces (all fields turning around) can occur
on any equipotential hyper-surface on dimension ranging  from zero to
$N-1$, and lying on the critical curve $B(\phi)  \propto\sqrt{-
  V(\phi)}$. Partial bounces can also occur away from the critical
curve, when only a subset of the fields inverts its flow direction.   

We have pointed out  that one {\em may} choose not to work with
gradient flows, for example if one wants to classify bulk solutions
according to some conserved charges. Rather than  the superpotential,
one is then led to use a non-separable Hamilton principal function  as
a generating function of a first order flow in the {\em full} parameter space of
the metric plus scalar fields. Although the case with global symmetries
is not relevant for holography, since there all symmetries must
necessarily be gauged, it can be useful to consider for cosmological
applications, something that we think would be worth exploring
further.  

This work may be extended in several ways. In holography, as we have
explained, all symmetries of the scalar manifold must be gauged, and
in order for  a solution with a non-zero charge one must turn on a
gauge field. This generically breaks Poincar\'e invariance, which
means a departure from a vacuum solution (e.g. by turning on a
chemical potential). It would be desirable to have a general first
order formalism in the presence of multiple scalars {\em and} gauge
fields, as it would be of interest both for condensed matter
applications and for holographic models of QCD. 

 Another way of obtaining  non-vacuum solutions is by going to finite
 temperature and considering black hole solutions. In  this context,
 already in the case of pure $AdS$ gravity it was shown how to write
 the flow equations in terms of a non-separable Hamilton principal
 function \cite{Lindgren:2015lia}. It would be interesting to work out the
 extension of to single- or multi-field black hole solutions. 

Finally, as we have mentioned above, many models of scalar field
inflation have multiple fields, and are intrinsically not reducible
to the single field case. It would be interesting to investigate
whether the formalism we have developed here can be used to describe
near-de Sitter solutions and whether using non-gradient flows associated
with conserved quantities can simplify the analysis  of
the space of solution.

\addcontentsline{toc}{section}{Acknowledgement}
\section*{Acknowledgement}
 
The authors would like to thank Elias Kiritsis, who participated in the early stages of this
work, for extended discussion and comments. This work was supported in part by the Advanced ERC grant SM-grav, No 669288.


\newpage
\appendix
\renewcommand{\theequation}{\thesection.\arabic{equation}}
\addcontentsline{toc}{section}{Appendices\label{app}}
\section*{Appendix}

\section{Non-gradient flows from a non-separable principal function}
\label{a:Non}

In this appendix the main equations we will need will be \eqref{lag7}
\eql{Non1}{
	e^{-dA}
	\le[
		{1\over 4d(d-1)}\le({\p\SS\over \p A}\ri)^2
		-\ha \GG^{rs}{\p\SS\over \p \phi^r}{\p\SS\over \p \phi^s}
	\ri]
	+e^{dA}V(\phi)
	=0 ~,
}
together with the following combination of \eqref{lag4} and \eqref{lag6}:
\begin{subequations}\label{Non2}
\begin{align}
	&\dot \phi^r
	=-e^{-dA}\GG^{rs}{\p\SS\over \p \phi^s}
	~,
	\label{Non2a}
	\\
	&\dot A
	=\frac{e^{-dA}}{2d(d-1)}{\p\SS\over \p A}
	~
	.
	\label{Non2b}
	\end{align}
\end{subequations}
One simple equation which follows immediately from the HJ \eqref{Non1} and will used below can be derived by first multiplying the HJ equation by $\exp(dA)$ and differentiating the resulting expression with respect to $A$:
\eql{Non3}{
0=\p_A
	\le[
		{\le(e^{-dA}\p_A\SS\ri)^2\over 4d(d-1)}
		-
		\ha e^{-2dA}\le({\p^r\SS}\p_r{\SS}\ri)
	\ri]
}
The first equation we can derive from \eqref{Non1} and \eqref{Non2} and which will be useful below is \eqref{gen3c}. We start by differentiating \eqref{Non2b} with respect to $u$
\begin{align}\label{Non4}
\ddot A
	&
	=\frac{1}{2d(d-1)}
	\le[
		\dot A \p_A\le(e^{-dA}{\p\SS\over \p A}\ri)
		+
		\dot \phi^r\p_r\le(e^{-dA}{\p\SS\over \p A}\ri)
	\ri]
	\nonumber
	\nonumber
	\\
	&=-\frac{e^{-2dA}}{2(d-1)}\le({\p^r\SS}\p_r{\SS}\ri)
	\nonumber
	\\
	&=-\frac{1}{2(d-1)}\GG_{rs}\dot \phi^r\dot \phi^s
\end{align}
where we used \eqref{Non3}.

Assume we have one solution $\SS(\phi,A)$ of equation \eqref{Non1}. From \eqref{Non2a} it is clear that in general the flows are not given by functions of the scalar fields alone. However, it is possible to {\it project} the flows on field space by the following procedure. We start by deriving the key element of this procedure, the function $\AA(\phi)$ defined over the scalar manifold such that it coincides with $A(u)$ on solutions:
\eql{Non5}{
A(u)=\AA(\phi(u))
}
The existence of $\AA(\phi)$ on a neighbourhood of any point where $\SS$ is locally single-valued is guaranteed provided that one can solve \eqref{Non2b} for $\AA(\phi)$ satisfying \eqref{Non5}:
\eql{Non6}{
e^{dA}\dot A=\GG^{rs}\le[\p_rS(\phi,A)\ri]_{A=\AA(\phi)}\p_s\AA={1\over 2d(d-1)}\p_A\SS(\phi,A)\big|_{A=\AA(\phi)}
}
The non-linear, first-order partial-differential equation on the scalar manifold  \eqref{Non6} is expected to include an integration function of $N-1$ variables.
Once a solution to \eqref{Non6} is given we can start locally projecting the solution associated with $\SS(\phi,A)$ on the scalar manifold.
As there will be many projected quantities, we define the following simplified notation:
\eql{Non7}{
\le.f(\phi,A)\ri|_{A=\AA(\phi)}\equiv \le.f\ri|_\AA
}
We define the projected functions 
$W(\phi)$, $\widehat W(\phi)$ 
and the projected vector field $\pi_r(\phi)$ through:
\begin{subequations}\label{Non8}
\begin{align}
&W(\phi):=-d^{-1}e^{-d\AA(\phi)}\le[\p_AS(\phi,A)\ri]_\AA~.
\label{Non8a}
\\
&\widehat W(\phi,A):=-e^{-dA}S(\phi,A)~,
\label{Non8b}
\\
&\pi_r(\phi):=\le[\p_r\widehat W\ri]_\AA=-e^{-d\AA(\phi)}\le[\p_rS(\phi,A)\ri]_\AA~.
\label{Non8c}
\end{align}
\end{subequations}

When equations \eqref{Non8} are combined with \eqref{Non2} we obtain:
\begin{subequations}\label{Non9}
\begin{align}
	&\dot \phi^r
	=\GG^{rs}\pi_s=\pi^r
	~,
	\label{Non9a}
	\\
	&\dot A
	=-\frac{W(\phi)}{2(d-1)}
	~
	.
	\label{Non9b}
	\end{align}
\end{subequations}
The HJ equation \eqref{Non1} assumes the same form as \eqref{v3} with $\pi_r$ playing the role of $\pi_r$:
\eql{Non10}{
\ha \pi_r \pi^r-{d\over 4(d-1)}W^2-V=0,
}
A complete equivalence of a generic non-separable principal function $\SS$ under the assumption \eqref{Non5} is shown if we derive from the Hamilton-Jacobi formalism equation \eqref{v5} with $\phi_r$ replaced by $\pi_r$. This is what we will do now.
We start rewriting \eqref{Non4} using the definitions \eqref{Non8}:
\eql{Non11}{
\pi^r\le(\pi_r-\p_rW\ri)=0~.
}
In other words, we can write $\pi_r$ as a sum of a gradient and a non-gradient part:
\eql{Non12}{
\pi_r=\p_rW+\xi_r \sp \pi^r\xi_r=0~.
}
With the definitions \eqref{Non8} we can explicitly write $\xi$ in terms of $\widehat W$ by a direct comparison of $\pi_r$ and $\p_r W$. It is convenient to first write $W(\phi)$ in terms of $\widehat W$ by substituting the definition \eqref{Non8b} into \eqref{Non8a}:
\eql{Non13}{
W(\phi)
	=
		\le[
			\widehat W(\phi,A)
			+d^{-1}\p_A \widehat W(\phi,A)
		\ri]_\AA
}
By deriving the expression \eqref{Non13} with respect to $\phi^r$ we obtain:
\begin{align}
	\p_{r}W(\phi)
	&
	=
		\le[
			\p_r\widehat W(\phi,A)
			+\p_A\widehat W(\phi,A)\p_r\AA
			+d^{-1}\le(
					\p_r\p_A \widehat W(\phi,A)
					+\p_A^2 \widehat W(\phi,A)\p_r\AA
				\ri)
		\ri]_\AA
	\nonumber
\\
	&
	=
		\le[
			\p_r\widehat W(\phi,A)
		\ri]_\AA
		+
		d^{-1}
		e^{-d\AA}
		\p_r\le[
			e^{dA}
			\p_A\widehat W(\phi,A)
		\ri]_\AA
\\
	&
	=
		\pi_r
		+
		d^{-1}
		e^{-d\AA}
		\p_r\le[
			e^{dA}
			\p_A\widehat W(\phi,A)
		\ri]_\AA
	\label{Non14}
\end{align}
where in the last line we used the definition \eqref{Non8c} of $\pi_r$ in terms of $\p_r\widehat W(\phi,A)$.
By comparing \eqref{Non12} and \eqref{Non14} we obtain the following two equivalent expressions for $\xi$:
\begin{align}
\xi_r=
	&-{1\over d}
		e^{-d\AA}
		\p_r\le[
			e^{dA}
			\p_A\widehat W(\phi,A)
		\ri]_\AA
	\nonumber
	\\
	&
	=
	-d^{-1}
	\le[
		\p_r\p_A \widehat W(\phi,A)
	\ri]_\AA
	-
	\le[
		\p_A\widehat W(\phi,A)
		+d^{-1}\p_A^2 \widehat W(\phi,A)
	\ri]_\AA\p_r\AA
\label{Non15}
\end{align}
Contracting \eqref{Non15} with $\pi^r$ gives back, after some algebra, equation \eqref{Non3}, showing that $\xi_r$ is indeed orthogonal to $\pi^r$.
To compute the curl of compute the curl of $\pi_r$ we notice that, from \eqref{Non12}, it equals the curl of $\xi_r$ and using \eqref{Non15}, we obtain:
\begin{align}
	\p_{[r}\xi_{s]}
	&
	=
	\p_{[r}W_{s]}
	=
	-\le[
		\p_{[r}\p_A\widehat W(\phi,A)
	\ri]_\AA\p_{s]}\AA
	=d~\xi_{[r}\p_{s]}\AA
\label{Non16}
\end{align}

We can contract \eqref{Non16} with $\pi^r$ and use \eqref{Non12} both to eliminate the term proportional to $\pi^r\xi_r$ and to rewrite the result solely in terms of $\pi_r$ and $\AA$. The results is the following expression:
\begin{align}
	\pi^r
	\le(
		\p_{r}W_{s}
		-\p_{s}W_{r}
	\ri)
	=
	-d~\le(\pi^r
	\p_{r}\AA\ri)
	\xi_{s}
	=
	-d~\le(\pi^r
	\p_{r}\AA\ri)
	\le(W_{s}-\p_sW\ri)
	\label{Non17}
\end{align}
Because $\AA$ is a solution to equation \eqref{Non6}, we can rewrite this equations in terms of the quantities defined in \eqref{Non8} leading to the following identity on field space:
\eql{Non18}{
\pi^r\p_r\AA=-{1\over 2(d-1)}W(\phi)~.
}
Substituting \eqref{Non18} into \eqref{Non17} leads to the expression:
\begin{align}
	\pi^r
	\le(
		\p_{r}W_{s}
		-\p_{s}W_{r}
	\ri)
	=
	\le({d\over 2(d-1)}W\ri)
	\le(W_{s}-\p_sW\ri)
	\label{Non19}
\end{align}
When $W$ is non-zero we can rearrange equation \eqref{Non19} in the following form:
\eql{Non20}{
\pi_s=\p_sW+{2(d-1)\over d~W}\pi^r\le(\p_r\pi_s-\p_s\pi_r\ri)~.
}
Equation \eqref{Non20} is nothing but equation \eqref{v5}, proving that when a solution $\SS$ to the HJ equations is non-separable and there is a function $\AA(\phi)$ satisfying (\ref{Non5}-\ref{Non6}), the flows can be projected on field space and result in a non-gradient velocity field for $\phi^r$.

The well known fact that when the principal function $\SS$ factorises, i.e., when $\widehat W$ defined on \eqref{Non8b}, is independent of $A$ follows here from the first line of \eqref{Non15} where we see that $\xi$ vanishes identically and from equation \eqref{Non12}.
Imposing that the flow is gradient, i.e., $\xi$ vanishes identically, tells us that $\widehat W$ {\it can} depend on $A$ but only in specific way that we show below.

\paragraph{A gradient flow given by $W(\phi)$ does not necessarily imply a separable principal function $\SS(\phi,A)$.}
From now on {\it assume that we know $W(\phi)$ for a curl-free velocity field $\pi^r$ but do not know $S(\phi,A)$ and we want to determine it}.
Equation \eqref{Non20} imply immediately that $\pi_r$ is the gradient of $W$, so the flows are gradient flows. From the definition of $\xi$ in equation \eqref{Non12} we know that it must vanish and, using the first line of \eqref{Non15} for finite $\AA$ this means:
\eql{Non21}{
0=
		\p_r\le[
			e^{dA}
			\p_A\widehat W(\phi,A)
		\ri]_\AA
		\implies
		\le[e^{dA}
			\p_A\widehat W(\phi,A)
		\ri]_\AA=C~ d
}
where the factor of $d$ is chosen for later convenience and $\AA$ is a solution of equation \eqref{Non18} which can be written for gradient flows as:
\eql{Non22}{
-2(d-1)\p^r\log(W)\p_r\AA=1.
}
We can write the most general solution to \eqref{Non22}  in terms of an arbitrary vector field $\eta^r(\phi)$ which contains the integration function which specifies a given $\AA(\phi)$ as follows:
\begin{subequations}\label{Non23}
\begin{align}
&
\AA(\phi)=
			\int (Y_r+\eta_r)d\phi^r,
			\label{Non23a}
\\
&
Y_r(\phi)\equiv-{W\over 2(d-1)}{\p_rW\over \p_sW\p^sW}~,
			\label{Non23b}
\\
&
\p_{[r}\le(Y_{s]}+\eta_{s]}\ri)=0~,
			\label{Non23c}
\\
&
Y^s\eta_s=0~.
			\label{Non23d}
\end{align}
\end{subequations}
Equation \eqref{Non23c} is integrability condition which is equivalent to the requirement of vanishing torsion in field space. Expression \eqref{Non23a} naturally incorporates the inhomogeneous solution to equation \eqref{Non22}. Equation \eqref{Non23c} should be solved for $\eta_s$ subject to the orthogonality constraint \eqref{Non23d} and a  choice of a field $\eta_s$ among all possible solutions corresponds to a choice of a homogeneous solution to \eqref{Non22}.

Let $g(\phi,A)$ be {\it any} regular function of the scalar field and the scale factor and $\AA(\phi)$ any function satisfying \eqref{Non23}. Equation \eqref{Non21} implies, at most, that
\eql{Non25}{
	\p_A\widehat W(\phi,A)
	=
	e^{-dA}\le[
		C
		+g(\phi,\AA(\phi))
		-g(\phi,A)
	\ri] d~.
}
Integration of \eqref{Non25} demands the choice of an arbitrary function of the scalar fields, here called $f(\phi)$:
\begin{align}
	\widehat W(\phi,A)
	=
	& -e^{-dA}\le[
		g(\phi,\AA(\phi))+C
	\ri] 
	-\int^A \dd B e^{-dB}g(\phi,B)~ d~+f(\phi).
\label{Non26}
\end{align}
Since we have started with the assumptions that we know $\pi_r$ which is closed, i.e. $\p_{[r}W_{s]}$ vanishes, and we know $W(\phi)$, we can use equation \eqref{Non13} to relate $W(\phi)$ and $f(\phi)$:
\begin{align}
	f(\phi)
	=
	W(\phi)
	+\int^\AA \dd B e^{-dB}g(\phi,B)~ d
	+e^{-d\AA}	g(\phi,\AA)
	\label{Non27}
\end{align}
Substitution of \eqref{Non27} into \eqref{Non26} yields:
\begin{align}
	\widehat W(\phi,A)
	=
	& -e^{-dA}C
	+W(\phi)
	+\int^\AA_A \dd B e^{-dB}\le[g(\phi,B)-g(\phi,\AA)\ri]~d.
	\label{Non28}
\end{align}

Substituting \eqref{Non27} into \eqref{Non26} we obtain:
Having the general expression \eqref{Non26} we can express Hamilton's principal function in terms of the unknown functions $g(\phi,A)$ and $f(\phi)$ by multiplying \eqref{Non25} by $-\exp(dA)$, as prescribed by \eqref{Non8b}.
\begin{align}
	\SS(\phi,A)
	=
	C
	-e^{dA}W(\phi)
	+\int^\AA_A \dd B e^{d(A-B)}\le[g(\phi,B)-g(\phi,\AA)\ri]~d.
	\label{Non29}
\end{align}
It can be convenient to relate the function $f(\phi)$ to the function $W(\phi)$ defined in \eqref{Non8a}.

For the special case in which $g(\phi,A)$ vanishes, substitution of \eqref{Non26} in equation \eqref{Non13} implies that $f(\phi)$ equals $W(\phi)$:
\eql{Non30}{
	g(\phi,A)\implies	\widehat W(\phi,A)=W(\phi)-Ce^{-dA}~.
}
The principal function can be reconstructed via \eqref{Non8b} and the result is:
\eql{Non31}{
\SS(\phi,A)=-e^{dA}W(\phi)+C\sim -e^{dA}W(\phi)~.
}
The last step means that a vanishing and a non-vanishing $C$ are equivalent as $\SS$ is defined up to an arbitrary additive constant. Therefore, the absence of a curl implies the existence of a $\SS$ which factorises as in \eqref{Non28}, as $\widehat W$ can be chosen from \eqref{Non27} with vanishing $C$, but other, non-separable principal functions exist which yield the same gradient flows.

\section{A non-gradient velocity field for two scalars: analytic example.}
\label{a:ex}

 We can build a non-gradient solution by first choosing the paths that the we want in such a way that the velocity field they generate has the desired profile. In this example we proceed with 
Consider two scalar fields on a flat field-space and we define polar coordinates on $\MM_\phi$:
\eql{sn0}{
\begin{cases}
\phi^1=R \cos \T
\\
\phi^2=R \sin \T
\end{cases}
}
so that the metric components are:
\eql{sn1}{\GG_{RR}=1,\quad\GG_{\T\T}=R^2,\quad\GG_{R\T}=0.}
We are interested in constructing paths which define Archimedean spirals:
\eql{sn2}{R=\a~(\T-\T_0),\qquad \T_0\in[0,2\pi)~.}

\begin{figure}[t]
\centering
\includegraphics[width=0.55\textwidth]{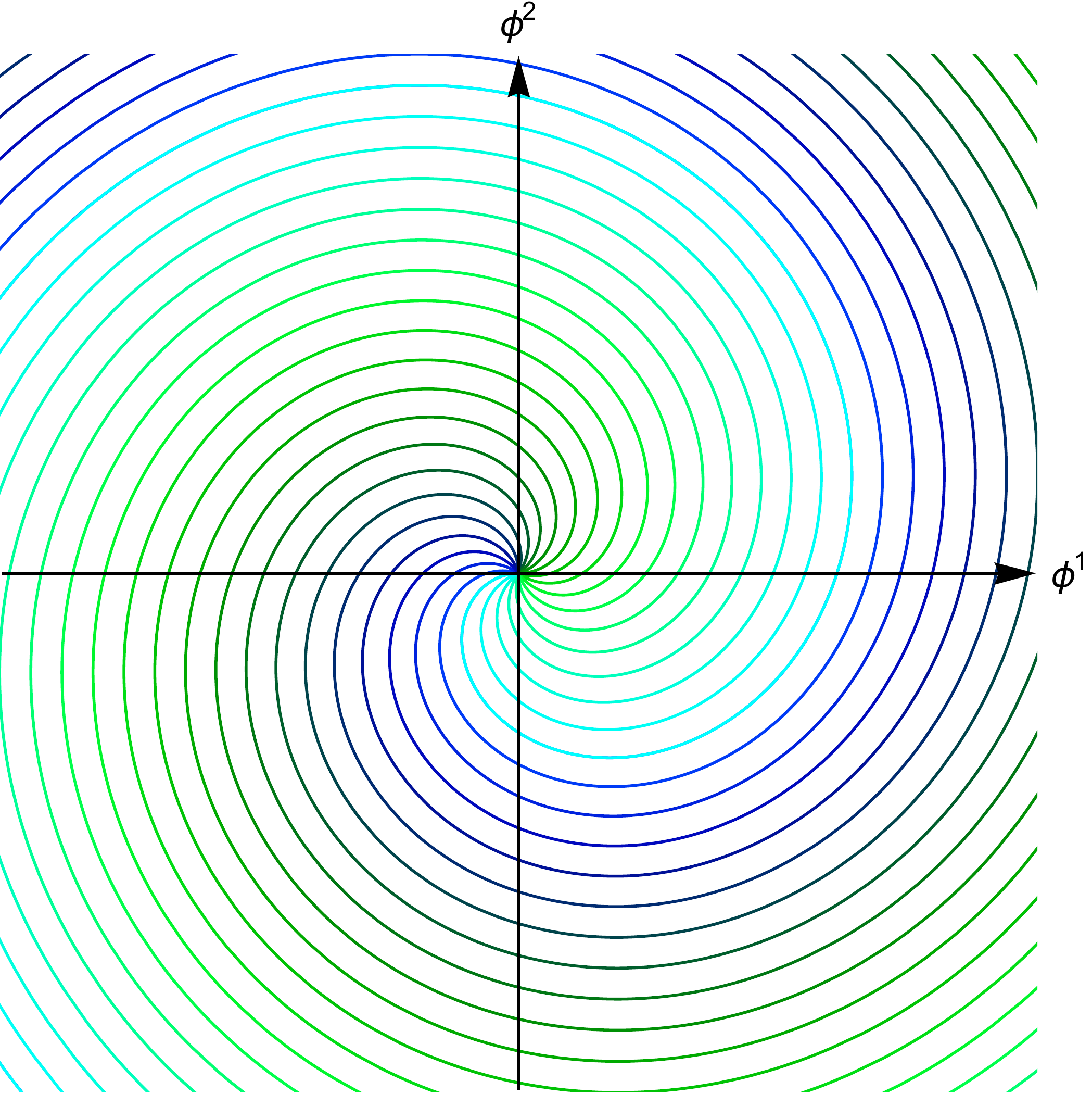}
\caption{Archimedean spirals \protect\eqref{sn2} with differerent values of $\T_0$ for $\a$ given by \protect\eqref{sn10}. The corresponding velocity field \protect\eqref{sn3} has a non-zero curl \protect\eqref{sn4}.} 
\label{f:sn1}
\end{figure}

Each angle in \eqref{sn2} defines a curve of the family, as shown in figure \ref{f:sn1} for a particular choice of $\a$. The corresponding velocity field can be written as:
\begin{subequations}\label{sn3}
\begin{align}
&\dot R(u)=W^R(R)\equiv R~\g(R),\label{sn3a}\\
&\dot \T(u)=W^\T(R):=\a~R~\g(R)~.\label{sn3b}
\end{align}
\end{subequations}
The function $\g(R)$ encodes the rate of change of $R$ through \eqref{sn3a} and the factor of $R$ relating $\g$ to $W^R$ is chosen for later convenience in order to make the power series expansion of $\g$ regular in $R$.
The curl of this velocity field is
\eql{sn4}{
\FF_{R\T}=\p_R \pi_\T=\a \le(R^3\g(R)\ri)'.
}
where we used the metric \eqref{sn1} to lower the $\T$ index. This velocity field must be accompanied by a function $W(\phi)$ which satisfies equation \eqref{v5}. We choose it to be rotationally invariant: $W=W(R)$. With this assumption and \eqref{sn4}, equation \eqref{v5} become the following pair of equations:
\begin{subequations} \label{sn5}
\begin{align}
	&W_R
	=
	\p_R W
	-
	{ 2(d-1)\over d~W} 
		W^\T\a \le(R^3\g(R)\ri)',
		 \label{sn5a}
	\\
	&\pi_\T
	=
	+
	{ 2(d-1)\over d~W} 
		W^R\a \le(R^3\g(R)\ri)'
		 \label{sn5b}
\end{align}
\end{subequations}
Equation \eqref{sn5b} can be seen as defining $W(R)$ in terms of the unknown function $\g$, as $W^R$ and $W^\T$ are related to it via \eqref{sn3}:
\begin{align}\label{sn6}
	&W(R)
	=
	+
	{ 2(d-1)\over d~R^2} 
		{\le(R^3\g(R)\ri)'}
\end{align}
With \eqref{sn6} providing the functional form of $W(R)$, equation \eqref{sn5a} provides a differential equation to be solved for $\g(R)$:
\eql{sn8}{
0	=
	\p_R \le(
	{{\p_R\le(R^3\g(R)\ri)}\over R^2} 
	\ri)
	-
	{ d\over 2(d-1)}\le( 1+\a^2 R^2 \ri)R\g(R).
}
Among the two linearly independent solutions of equation \eqref{sn8} the one which provides a regular velocity field at the origin is
\begin{align}\label{sn9}
\g(R)=
&\frac{\left(2 \sqrt[4]{2} \Gamma \left(\frac{3}{2}-\frac{\alpha +\sqrt{\frac{d}{2 d-2}}}{4 \alpha }\right)\right) \left(e^{\frac{1}{2} \alpha  \sqrt{\frac{d}{2 d-2}} r^2} U\left(-\frac{\alpha +\sqrt{\frac{d}{2 d-2}}}{4 \alpha },-\frac{1}{2},-\sqrt{\frac{d}{2 d-2}} r^2 \alpha \right)\right)}{\sqrt{\pi } \left(\sqrt[4]{2} r^3\right)}+
\nonumber
\\
&
~~
-\frac{\sqrt[4]{2} \left(e^{\frac{1}{2} \alpha  \sqrt{\frac{d}{2 d-2}} r^2} L_{\frac{\alpha +\sqrt{\frac{d}{2 d-2}}}{4 \alpha }}^{-\frac{3}{2}}\left(\alpha  \left(-\sqrt{\frac{d}{2 d-2}}\right) r^2\right)\right)}{\left(\sqrt[4]{2} r^3\right) L_{\frac{\alpha +\sqrt{\frac{d}{2 d-2}}}{4 \alpha }}^{-\frac{3}{2}}(0)}
\end{align}
However, when $\a$ has the following value:
\eql{sn10}{
\a=-\sqrt{d\over 2(d-1)}
}
the regular solution \eqref{sn9} vanishes identically and the regular solution to \eqref{sn8} is given by a simple expression in integral form:
\begin{align}\label{sn11}
\g(R):=&\frac{(d-1)}{\ell R^2} e^{\frac{d }{4 (d-1)}R^2} \left(1-\int_0^1 \exp \left(\frac{d \left(x^2-1\right)}{2 (d-1)}R^2\right) \, dx\right)
\end{align}
For the choice \eqref{sn10}, it is simple to substitute $\g$ from \eqref{sn11} into \eqref{sn3} and \eqref{sn6} in order to obtain $\pi^r$ and $W$, the later simplifying to:
\be
W(R)=2(d-1)\frac{1}{\ell}\exp(\frac{d R^2}{4 (d-1)})-R^2 \g (R)
\label{sn12}
\ee
With these quantities, we can construct the potential $V(R)$ that has these flows as solutions, by the use of the algebraic equation \eqref{v3}. The result is:
\begin{align}
V(R)
=
&
\ha\le(\g (R)R\ri)^2
-\frac{ d}{ \ell}\le(
	{(d-1) \over  \ell}
	-\g (R)R^2
 \ri)
\exp({d\over2 (d-1)}R^2)
\label{sn13}
\end{align}
where the function $\g(R)$ is given by \eqref{sn11}. The first terms of the series expansion of the potential \eqref{sn13} around the origin are:
\eql{sn14}{
V(R)=
-\frac{(d-1) d}{l^2}-\frac{d^2 R^2}{9 l^2}-\frac{7 d^3 R^4}{360 \left((d-1) l^2\right)}+O\left(R^5\right)
}

Equation \eqref{floA} with $W$ given by \eqref{sn12} leads to a scale factor that diverges as an asymptotically AdS warp factor as $R$ tends to zero. Through the holographic dictionary this would naively means that the we should associate a UV fixed point at $R=0$ for the flows solving \eqref{sn2}.
There is of a curl, as follows from \eqref{curl10}, but in the $R$ expansion the curl is sub-leading close to the origin. In cartesian coordinates:
\begin{subequations}\label{sn15}
\begin{align}
&\dot \phi^1(u)={d\over 3}\phi^1(u)+\cO(\phi)^2,\\
&\dot \phi^2(u)={d\over 3}\phi^2(u)+\cO(\phi)^2.
\end{align}
\end{subequations}
With \eqref{sn11} the curl acquires a simpler form in terms of $\g$ and its first derivative:
\eql{snF1}{
\FF_{R\T}=
-\sqrt{d\over 2(d-1)}R^2\le[3\g(R)+R\g'(R)\ri].
}
is therefore a sub-leading property of the flows close to the UV, showing that for these solutions the first term on the right-hand side of equation \eqref{recall} dominates over the second. In other words, \eqref{recall} mixes leading and sub-leading terms of the $R$ expansion. We will see now that if we try to interpret this in holographic way, this mixing amounts to {\it fix a relation between sources and VEVs} at the UV and not by fixing the source. 

\section{Local reconstruction of the superpotential} \label{app:local}

In this appendix we show how to locally reconstruct a
superpotential, given a solution. The idea  is that, around a generic
point along a given flow  $(A(u), \phi^r(u))$ we can make an
appropriate   coordinate transformation in field space from $\{\phi_r\} \to (\xi,
\eta_\alpha)$ with only $\xi$ changing along the flow and
 $N-1$ ``spectator'' coordinates $\eta_\alpha$, which parametrise
the directions orthogonal to the flow. In these coordinates one can
reduce locally to a single-field flow, which therefore admits a local
superpotential  $W_{loc}(\xi)$, independent of $\eta_\alpha$,  for which the flow equation simply become
\be \label{loc1}
\dot{A} = -2(d-1)W_{loc}, \quad  \dot{\xi} =  W_{loc}', \quad \dot{\eta_\alpha} = 0.
\ee 
Below we show how this construction works explicitly at lowest order
in the $u$-dependence of the solution. 

We  from a solution of equations (\ref{gen3}), and we expand it in $u$
around a point $u_0$ where the gradient vector $\dot{\phi_r}$ is
non-vanishing. Without loss of generality we can set $u_0=0$ and $A(0)
= \phi_r(0) = 0$ by coordinate transformations and field
redefinitions. To lowest order around $u=0$ we can write: 
\be\label{loc2}
A(u) = A_1 u + {1\over 2} A_2 u^2 + O(u^3), \quad \phi^r(u) = \bar{\phi}^r u + O(u^2), 
\ee
where $A_1,A_2$ and $\bar{\phi}^r$ are constants and by assumption not all the
$\bar{\phi}^r$ vanish. We need to consider  the term $A_2$ because it enters
in equation (\ref{gen3c}) at the same order as  $\dot{\phi}^2$, which
starts as a constant as $u\to 0$.   Using
equations (\ref{gen3b}-\ref{gen3c}) we can determine  $A_1$ and $A_2$
to be 
\be \label{loc3}
A_1 =  -{1\over 2(d-1)} W_1, \qquad  A_2 = -{1\over 2(d-1)} \GG_{rs}\bar{\phi}^r\bar{\phi}^s
\ee
where $W_1$ is determined algebraically by the equation:
\be \label{loc4}
{d \over 4(d-1)} W_1^2 - {1\over 2}\GG_{rs}\bar{\phi}^r\bar{\phi}^s  =
  -V(0).  
\ee
The velocities $\bar{\phi}^r$ are then determined by equation
(\ref{gen3a}) from the derivatives of the potential and of the metric.  

We now define the new field variable  $\xi$ by
\be\label{loc5}
\xi  = \left(\GG_{rs}\phi^r\phi^s\right)^{1/2}.  
\ee 
Along the flow, around $u=0$ it behaves as
\be\label{loc6}
\xi(u) = \bar{\xi} u + O(u^2),  \quad \bar{\xi} \equiv
\left(\GG_{rs}\bar{\phi}^r\bar{\phi}^s\right)^{1/2} \neq 0.  
\ee
We can in principle define a coordinate transformation around  $\xi=0$
by adding $N-1$ coordinates $\eta_\alpha$ orthogonal to the flow,
which to this order will be independent of $u$ along the solution. 
 
We now {\em define} the local superpotential around the point $\xi = 0$ by:
\be \label{loc7}
W(\xi) = W_1 + \bar{\xi} \xi + O(\xi^2),
\qquad  
\ee
where $W_1$ is the same constant determined from equation
(\ref{loc4}).  
It is now straightforward  to check that (\ref{loc7}) serves, locally,
as a  superpotential for the flow close to $u=0$, 
\be
\dot{\phi^r} = \GG^{rs} \de_s W_{loc}, \qquad \dot{A} = -2(d-1) W_{loc} 
\ee
up to terms of order $u$. This construction works as along as
$\bar{\xi} = 0$, i.e. if the starting point around which we expand is
{\em not} a complete bounce. 

\section{Non-separable solution with angular momentum} \label{a:nonsep}
We consider the two-field case, with radial and angular variables $R$
and $\Theta$,   diagonal and rotationally invariant metric
$(\GG_{RR}(R),\GG_{\Theta\Theta}(R) )$ and  rotationally invariant
potential $V = V(R)$. We start from HJ equation (\ref{lag7}) and look for solution of the
form (\ref{curl2}), 
\be\label{L0}
\SS(A,R,\Theta) = \SS_0(A,R) + L \Theta, 
\ee
for which   equation (\ref{lag7}) becomes,
\be\label{L2}
e^{2dA}V(R) + \left[{1\over 4d(d-1)} \left( \de \SS_0 \over \de
      A\right)^2 - {1\over 2}\GG^{RR} \left( \de \SS_0 \over \de
      R\right)^2 -   {L^2\over 2}\GG^{\Theta\Theta} \right].
\ee  
We now look for solutions in an expansion in large $e^{dA}$, 
\be \label{L1}
\SS_0(A,R) = -e^{dA} W_0(R) + W_1(R) + e^{-dA}W_2(R) +\ldots. 
\ee
Inserting this ansatz in equation (\ref{L2}) we find: 
\begin{itemize}
\item {\bf Order $e^{2dA}$:}
\be\label{L3}
{d\over 4(d-1)} W_0^2 -  {1\over 2}\GG^{RR} \le({dW_0 \over dR}\ri)^2
+ V = 0
\ee 
i.e. the ``reduced'' superpotential equation with respect depending
only on $R$; 
\item {\bf Order $e^{dA}$:}
\be\label{L3b}
\GG^{RR}{dW_0 \over dR}{dW_1 \over dR} = 0
\ee
which implies constant $W_1 = C_1$; 
\item {\bf Order $e^0$:}

\be\label{L4}
{d\over 2d(d-1)} W_0 W_2 +  \GG^{RR}{dW_0 \over dR}{dW_2 \over dR} =  {L^2\over 2}\GG^{\Theta\Theta}
\ee
This is the lowest  order at which $L$ enters in $\SS_0$ (notice that
it enters in  $\SS$ in equation (\ref{L0}) at one order higher through
the $L \Theta$ term). The solution of equation (\ref{L4}) is
\be\label{L5}
W_2 =\exp\left(-{d\over 2(d-1)} \int {W_0 \over \GG^{RR} W_0' }
\right) \le[C_2 + L^2 \int {\GG^{\Theta\Theta}  \over \GG^{RR} W_0'}
\exp\left({d\over 2(d-1)} \int {W_0 \over \GG^{RR} W_0' }\right) \ri]
\ee   
where $C_2$ is another integration constant. 
\end{itemize}

\newpage

\addcontentsline{toc}{section}{References}
\bibliography{Multifield.bib}
\bibliographystyle{JHEP}
\end{document}